\documentclass[10pt]{article}
\usepackage[ansinew]{inputenc}
\usepackage{graphics}

\usepackage{latexsym,bezier,enumerate,longtable,dcolumn,color,pstricks,xspace,curves,mathpazo,url,xcolor}
\usepackage{amsmath,amsthm,amsopn,amstext,amscd,amsfonts,amssymb,fancybox,pst-node,pst-tree}
\usepackage{multicol}

\usepackage{epsfig,epic,graphicx,pstricks}

\usepackage{tikz}
\usetikzlibrary{positioning}

\mathchardef\isinpunto="0010

\newcommand{\R}{{\cal R}}

\newcommand{\D}{{\cal D}}

\newcommand{\T}{{\cal T}}
\newcommand{\K}{{\cal K}}

\newcommand{\abs}[1]{\lvert#1\rvert}

\newcommand{\G}{{\Gamma}}

\newcommand{\tq}{\,\vert\,}

\theoremstyle{plain}
\newtheorem{proposition}{\bf Proposition}
\newtheorem{theorem}{\bf Theorem}
\newtheorem{lemma}{\bf Lemma}
\newtheorem{corollary}{\bf Corollary}
\theoremstyle{definition}

\newtheorem{example}{\bf Example}

\title{\baselineskip .2in
\bf The position value as a centrality measure in social networks\footnote{This research has been
supported by I+D+i research project  PID2020-116884GB-I00 from the Government of Spain.}}
\date{}

\begin{document}
\setlength{\abovedisplayskip}{5mm} 
\setlength{\abovedisplayshortskip}{3mm} 
\setlength{\belowdisplayskip}{6mm} 
\setlength{\belowdisplayshortskip}{5mm} 

\maketitle \vspace*{-1cm} {\baselineskip .2in}
\begin{center}
{\small
\begin{tabular}{l} \bf Susana L\'opez
\\ Dpt. An\'alisis Econ\'omico y Econom\'{\i}a Cuantitativa, Universidad Aut\'onoma de Madrid,  Spain
\\ e-mail: \url{susana.lopez@uam.es}
\\[2mm]
\bf Elisenda Molina
\\ Instituto de Matem\'atica Interdisciplinar (IMI), Dpt. Estad\'{\i}stica e Investigaci\'on Operativa,
\\ Universidad Complutense de Madrid, Spain
\\ e-mail: \url{elisenda.molina@ucm.es}
\\[2mm]
\bf Martha Saboy\'a
\\ Dpt. An\'alisis Econ\'omico y Econom\'{\i}a Cuantitativa, Universidad Aut\'onoma de Madrid,  Spain
\\ e-mail: \url{martha.saboya@uam.es}
\\[2mm]
\bf Juan Tejada
\\ Instituto de Matem\'atica Interdisciplinar (IMI), Dpt. Estad\'{\i}stica e Investigaci\'on Operativa,
\\ Universidad Complutense de Madrid, Spain
\\  e-mail: \url{jtejada@mat.ucm.es}
\end{tabular}}
\end{center}

\begin{abstract}
The position value, introduced by Meessen (1988), is a solution concept for cooperative games in which the value assigned to a player depends on the value of the connections or links he has with other players. This concept has been studied by Borm {\it et al.} (1992) and characterised by Slikker (2005). In this paper, we analyse the position value from the point of view of the typical properties of a measure of centrality in a social network. We extend the analysis already developed in G\'omez {\it et al.} (2003) for the Myerson centrality measure, where the symmetric effect on the centralities of the end nodes of an added or removed edge is a fundamental part of its characterisation. However, the Position centrality measure, unlike the Myerson centrality measure, responds in a more versatile way to such addition or elimination. After studying the aforementioned properties, we will focus on the analysis and characterisation of the Position attachment centrality given by the position value when the underlying game is the attachment game. Some comparisons are made with the attachment centrality introduced by Skibski {\it et al.} (2019).

\end{abstract}

\section{Introduction}
\label{Introduction}

The important role of social networks nowadays is irrefutable, as evidenced by their emergence in a large variety of fields of application. Social networks play a fundamental role in sociology, describing the existing relations among the members of a society. They allow to detect key members (Borgatti, 2003), communities (Girvan and Newman, 2002), to analyse rumour spreading (Zubiaga {\it et al.}, 2018), etc. But, social networks are no longer limited to the pure sociological range, they are also currently used in other fields which, at first sight, seem far removed. Among many areas of application, we can mention wildfire spreading (Hajian {\it et al.}, 2016), the analysis of the genome and proteome of a living being (Horvath, 2011; Estrada, 2006), as well as the analysis of systemic problems of banking contagion (Gofman, 2017).  


Formally, a social network is often modeled as a graph, with  nodes representing agents and  edges or arcs representing communication channels or relationships between them. This structure can be enriched with additional information such as the degree of relationship or influence between agents, or the weight of each agent, etc.

One of the most important tasks in the analysis of social networks is the evaluation of which nodes or agents are more important, relevant, powerful or influential. This evaluation can be used to select the main agents on which to focus our attention to explain or condition the network dynamics. To do this, the importance of each agent needs to be assessed, or at least the agents need to be ranked according to their importance, based on the notion of centrality. However, although the centrality of a node or agent in a social network is an intuitive concept, it is not unambiguously defined, it is multifaceted, since it depends on what exactly is being measured. 

Attempting to answer the question of what all centrality measures must have in common is difficult. As early as 1966, Sabidussi considered a set of criteria that a proper centrality measure should satisfy, but his criteria ``eliminate most known measures of centrality'' and ''they do not actually attempt to explain what centrality is'' (see Borgatti and Everett, 2006). 

The first idea to specify the meaning of centrality came from considering the position of the node compared to others in simple networks. For example, in a star graph, it seems that the most important node is the hub, because it is the node with the highest degree, it is the closest node to the rest of the nodes, and it is essential to establish the connection between two different nodes. Freeman (1979) reduced the existing centrality measures to these three basic concepts and provided three basic centrality measures: Degree, Closeness and Betweenness to capture each of them. These centrality measures reach their maximum values for the hub of a star. This property is considered to be the defining characteristic of true centrality measures. Subsequently, other centrality measures have been introduced. Borgatti and Everett (2006) extended Freeman's work. They conducted a graph-theoretic review of centrality measures and classified them along four key dimensions. Mirroring Freeman's classification, they also considered {\em degree-like} measures (e.g. Katz, 1953; Bonacich, 1987), {\em closeness-like} measures (e.g. Friedkin's {\it immediate effects centrality}, 1991), and {\em betweenness-like} measures (e.g. Newman, 2005).  Apart from these contributions, other attempts have been made to propose necessary properties or even axiomatic characterisations of a centrality measure, such as Garg (2009), Landherr {\it et al.} (2010), Boldi and Vigna (2014), and Bandyopadhyay {\it et al.} (2017). Recently, Bloch {\it et al.} (2021) have proposed a different taxonomy of centrality measures based on different statistics associated with each node that capture its position in the network. In conclusion, apart from some basic properties, there is no general agreement on what properties should characterise a centrality measure.

The analysis of centrality has also been approached from the theory of cooperative games.  Tarkowski {\it et al.} (2017) provide a survey of the main game-theoretic approaches to measuring centrality in social networks.

The basic idea is to understand the social network as a joint project in which different agents with defined interests cooperate in order to obtain benefits from their interaction with other agents in the network. Within this approach, game theory has enriched the existing measures by explicitly taking into account the functionality or different purposes that the social network can have. Some of these centrality measures integrate the topological information provided by the network with a coalitional game that models the interests or benefits that the nodes (players) can obtain as a result of their interactions (see, for example, Grofman and Owen (1982), G\'omez {\it et al.}, 2003, del Pozo {\it et al.}, 2011). Other approaches define a cooperative game directly from the network (Narayanam and Narahari (2010), Lindelauf {\it et al.} (2013), Musegaas {\it et al.} (2016)). In any case, usually, but not always, the game-theoretic centrality measure is obtained as the Shapley value (Shapley, 1953) of the proposed coalitional game.

Moreover, studying the centrality of nodes in a social network based on a value of an associated game allows us to use the properties of the game-theoretic value to derive properties of the centrality measure. For example, in G\'omez {\it et al.} (2003),  desiderata of key centrality properties is given and checked for the Myerson value based family of centrality measures they introduce. Following this approach, Skibski {\it et al.} (2018) and Skibski {\it et al.} (2019) give axiomatic characterisations of some game-theoretic centrality measures.

In this paper we propose to explore the possibilities of using the position value (Messens, 1988) to define a family of centrality measures. In this approach, the centrality measure of a node depends on the importance attributed to its links. This family, under certain conditions, satisfies typical properties of a centrality measure that have already been demonstrated for the Myerson centrality family. However, unlike Myerson's centrality, Position centrality escapes from the property of fairness to account for different effects of the addition of an edge over its end nodes according to their role in the previous social network.

Both game-theoretic centrality measures, the Myerson and Position centrality families, are complex measures that account for more than one of the key dimensions considered by Borgatti and Everett (2006). Specifically, Myerson and Position centralities can be rewritten to obtain more information about their behaviour as a centrality measure by decomposing their values into two parts that measure two different abilities of each member in the social network: its ability to connect and its ability to intermediate\footnote{In fact, Manuel {\it et al.} (2020, 2022) analysed each of these two parts separately, which they called the {\em within-group Position value (WG-Myerson, WG-Position values)} and the {\em between-group Position value (BG-Myerson, BG-Position values)}, from a game-theoretic point of view. They obtained axiomatisations for all Position and Myerson involved values.} (see G\'omez {\it et al.} (2003) and G\'omez {\it et al.} (2004)). That is, both measures are simultaneously {\em radial} and {\em medial} in terms of nodal involvement in the walk structure of the graph. However, they differ in the properties of the walks they measure. Myerson centrality only counts walks that are minimal with respect to nodes, whereas Position centrality only counts walks that are minimal with respect to edges. Moreover, each of these walks can be properly evaluated according to the goal achieved by the members they connect, by considering a game that captures the functionality of the social network. 

In addition to the study of the general properties of the family of Position centralities, we consider in particular the Position attachment centrality. For this measure, we study some specific properties and also characterised it following the work of Skibski et al. (2019) on Myerson attachment centrality.

The remainder of this paper is structured as follows. In Section 1, we review related work. In Section 2, we formally define some aspects of cooperative games, graphs and Myerson centrality. In Section 3, we propose Position centrality and some of its properties as a centrality measure. In Section 4, we present some results on the behaviour of the Position centrality measure when a new edge is added to the graph.
The special case of Position attachment centrality is analysed and characterised in Section 5. Finally, the last section concludes the paper. We have also included an appendix with the proof of some results in order to facilitate the reading of the paper.

\section{Preliminaries}
\label{Preliminaries}

First, we summarise the basic elements concerning the both topics we deal with, coalitional games and graphs. Then, we review the concept of {\em Myerson centrality} (Owen, 1986; G\'{o}mez {\it et al.}, 2003).


\subsection{Coalitional games}

A \emph{coalitional game} with transferable utility (TU-game) is a
pair $\left( N,v\right)$, where $N=\{1,2,3,...,n\}$\ is the set of players
and $v$, the\emph{\ characteristic function}, is a map $v:2^{N}\rightarrow 
\mathbb{R}$, with $v\left( \varnothing \right) =0$. For each coalition, 
$S\subseteq N$, $v(S)$ represents the transferable utility that $S$ can guarantee to 
obtain whenever its members cooperate. Let $G^{N}$ be the class of all coalitional games over player set $N$.
A TU-game $(N,v)$ is {\it superadditive}  if $v( S\cup T) \geq v\left( S\right) +v\left( T\right)$,  for all $S, T\subseteq N, S\cap T=\emptyset$, and it is {\it convex} if $v(S\cup T)+v(S\cap T)\geq v(S)+v(T)$,  for all $S,T\subseteq N$.  A TU-game $(N,v)$ is {\it symmetric} if for all $S\subseteq N$, $v\left( S\right) =f\left(s\right)$, where $s$ represents the cardinality of $S$.
It is {\it zero-normalized} if $v(\{i\}) = 0$,  for all $i\in N$. In this paper we mainly consider zero-normalized symmetric and superadditive games. We will denote by 
$G^N_0$ the subclass of these games in $G^N$.

For each $S\in 2^{N}\backslash \varnothing \ $, the \emph{unanimity game} $u_{S}\in G^{N}$ is defined by
\begin{equation*}
u_{S}\left( T\right) =\left\{ 
\begin{array}{cl}
1 & \text{if }S\subseteq T\text{,} \\ 
0 & \text{otherwise.}%
\end{array}%
\right.
\end{equation*}%
It is well-known that 
$\left\{ u_{S}\left\vert S\in 2^{N}\backslash \varnothing \right. \right\} $ is a basis of the vector space $G^{N}$, and each $v\in G^{N} $ can be expressed as follows:%
\begin{equation*}
v=\sum_{S\in2^{N}\backslash\varnothing}\Delta _{v}\left(
S\right) u_{S}\text{,}
\end{equation*}%
where $\left\{ \Delta _{S} (v)\right\} _{ \varnothing
 \neq S\subseteq N}$ is the set of the Harsanyi dividends of $v$ (Harsanyi, 1959), which are given by%
\begin{equation}
\Delta _{S}(v) =\sum_{T\subseteq S}\left( -1\right) ^{s -t}v\left( T\right) \text{,}
\label{harsanyi-div-def}
\end{equation}%
being $s=\left\vert S\right\vert $ and  $t=\left\vert T\right\vert $. 


A {\em value} $\varphi$ for TU games is a function that assigns to each game $(N,v)\in G^N$ a vector $\varphi(N,v)\in \mathbb{R}^N$, where $\varphi_i(N,v)\in \mathbb{R}$ represents the {\em value} of player $i$, $i\in N$. Shapley (1953) defines his value as follows:
\begin{equation}
\displaystyle
 \phi_i (N,v) = \sum_{\substack{S\subseteq N\\ i \notin
S}} \frac{s!(n-s-1)!}{n!} \bigl( v(S\cup \{ i\}) -
v(S)
 \bigr ), \quad i\in N.
\label{shapley-marginal}
\end{equation}
An alternative expression for the Shapley value in terms of the Harsanyi dividends is:%
\begin{equation}
\phi _{i}\left( N,v\right) =\sum_{i\in S\in 2^{N}\backslash \
\varnothing \ }\frac{\Delta _{S}(v) }{s }\text{.}
\label{shapley-dividends}
\end{equation}

\subsection{Graphs}

Let $\Gamma = (N, E)$, be an undirected graph without loops , where $N$ is the set of $n$ nodes and $E$ is the set of edges, $E\subseteq \{\{i,j\}/ i, j\in N, i\neq j\}$. Let ${\cal G}^N$ denotes the class of all undirected graphs without loops or parallel arcs with node set $N$. For all $i\in N$ we denote $E_i:=\{ e\in E \,\vert\, i\in e\}$, $E[S]:=\{ e\in E \,\vert\, e\subseteq S\}$, for all $S\subseteq N$, and $N[L]=\{ i\in N \tq i\in \bigcup_{e\in L} e\}$, for all $L\subseteq E$. The {\em subgraph induced by $S\subseteq N$} is  $\Gamma_S:=(S,E[S])$, whereas the {\em subgraph induced by $L\subseteq E$} is $\Gamma_L:=(N[L],L)$.

A graph is {\em connected} if every pair $i, j\in N$ of its nodes is connected directly or indirectly, i.e., if there is a path in the graph from node $i$ to node $j$; otherwise, the graph is {\em not connected}. The  relation of connectivity induces a partition of the
node set $N$ into {\em connected components}, two nodes being in the same
connected component if they are connected. Let us denote the collection of connected components of the graph $\Gamma$ by $\K(\Gamma)$. Moreover, for the sake of simplicity, $\K(S)$ denotes the collection of connected components of the graph $\Gamma_S$, and analogously 
$\K(L)$ denotes the collection of connected components of the graph $\Gamma_L$, for all $S\subseteq N$, and all $L\subseteq E$, respectively. For every node $i\in N$, $\K_i(\Gamma)\in \K(\Gamma) $, denotes the connected component of the graph $\Gamma$ to which node $i$ belongs. Analogously, if $i\in S$, or $E_i\cap L\neq  \emptyset$, $\K_i(S)\in \K(S)$, and $\K_i(L)\in \K(L)$ denote, respectively, the connected component of the corresponding graph to which node $i$ belongs. 
A cutvertex is a vertex whose removal increases the number of connected components, and a cutedge is an edge whose end nodes are both cutvertices. Let $\mathcal{D}\left( L\right)$ denote the set of cutedges of $L$.

\subsection{Myerson centrality}


As we have mentioned before, since 
the pioneer contribution of Grofman and Owen in 1982, several approaches to measure centrality of nodes in a social network that rely on cooperative TU games have been proposed in the literature. The goal of game theoretic centrality measures is to take into account the purpose of the social network. To be specific, considering an appropriate TU game $(N,v)$ to describe the {\em functionality} (i.e. the purpose) of the social network (to send messages, to develop a joint project, etc.), the {\em Myerson centrality} is defined as the Myerson value (1977) of the game $(N,v)$ taking into account the cooperation structure given by the social network $\Gamma=(N,E)$, that integrates both kinds of information: the structure and the goal of relations.

We follow the Myerson centrality approach, i.e., to define a  family of {\em game-theoretic} centrality measures by means of integrating the information about the benefit of each group $S\subseteq N$ given by the game with the restrictions in the communication between players modelled by the graph $(N,E)$, and selecting a particular value  for games with restricted cooperation as a centrality measure. In this context, the triplet $(N,v,E)$ is called a {\em communication situation}.

Formally, a {\em centrality measure} for social networks with symmetric relations is an assignation which associates to each graph $\Gamma=(N,E)\in {\cal G}^N$ a vector $\sigma(\Gamma)\in \mathbb{R}^N$, where $\sigma_{i}(\Gamma)\in \mathbb{R}$ represents the {\em centrality} of node $i$, $i\in N$.

Owen (1986) and G\'{o}mez {\it et al.} (2003) introduce a family of centrality measures that arises of considering {\em symmetric} TU games to describe the functionality, and relying on the Myerson value (1977) of the graph restricted game.


That is, given a symmetric game $(N,v)$, the {\em Myerson centrality} of nodes in the social network $\Gamma=(N,E)$ is measured by their Shapley values in the {\em graph-restricted game} $(N,v_E)$ (Myerson, 1977) that represents the economic possibilities of the agents when the available communication possibilities are taken into account:
$$
v_E (S)= \sum_{T\in \K (S)} v(T), \quad S\subseteq N.
$$
Thus, $\sigma_i(\Gamma):=\mu_i(N,v,E):=\phi_i(N,v_E)$, for all node $i\in N$, where $\mu(N,v,E)$ denotes the Myerson value of the game $(N,v)$ given the cooperation structure $\Gamma=(N,E)$.

%
For instance, Owen (1986) and G\'{o}mez {\it et al.}  (2003) consider some TU games modeling interesting functionalities such as sending messages between pairs of individuals ({\em messages game}), to develop a joint project ({\em overhead game}), or the ability to form groups of two or more individuals ({\em conferences game}).

\begin{itemize}
    \item {\em Messages game}: $(N,v_m)$, that counts the number of messages 
    that can be sent between pairs of individuals (in both ways), $v_m(S)=s(s-1)$, for all $\emptyset \neq  S\subseteq N$.  
    
    \item {\em Overhead game on $N$}: $(N,v_o)$, that accounts for the general cost that any set of players should pay to perform an action,   
    with $v_o(S)=-1$, for all $\emptyset \neq  S\subseteq N$. 
    
    \item {\em Attachment game}: In this paper, we will consider the attachment game $(N,v_a)$, given by $v_a(S)=2(s-1)$, for all $\emptyset \neq  S\subseteq N$, and introduced in  Skibski {\it et al.} (2019), which is proportional to the zero-normalisation of the overhead game.\footnote{The zero-normalisation of the overhead game , $v_o^0(S)=s-1$, is called {\em pure overhead game on $N$} in Owen (1986).} 
    
    \item {\em Attachment-Messages game}: $(N,v_{am})$; $v_{am}(S)=s^2+s-2$, for all $\emptyset \neq  S\subseteq N$.
    
  \item {\em Conferences game}: $(N,v_c)$, that counts the number of subsets in $S$ with cardinal greater than 1, and thus $v_c(S)=2^s-s-1$, for all $S\subseteq N$.
\end{itemize}

\begin{example}
The following example shows the impact of the functionality of the social network on the centralities of each node. 


\begin{figure}[h]
\begin{center}

\begin{tikzpicture}[node_style/.style={draw,circle,minimum size=0.5cm,inner sep=1}]

\node[node_style] (siete) at (0,1.2) {7} ;
\node[node_style] (seis) at (0.75,2.4) {6} ;
\node[node_style] (ocho) at (0.75,0) {8} ;
\node[node_style] (cuatro) at (1.5,1.2) {4} ;
\node[node_style] (cinco) at (2.25,2.4) {5} ;
\node[node_style] (nueve) at (2.25,0) {9} ;
\node[node_style] (uno) at (3,1.2) {1} ;
\node[node_style] (dos) at (4.5,1.2) {2} ;
\node[node_style] (tres) at (6,1.2) {3} ;
\node[node_style] (doce) at (9,1.2) {12} ;
\node[node_style] (once) at (8.25,2.4) {11} ;
\node[node_style] (trece) at (8.25,0) {13} ;
\node[node_style] (quince) at (7.5,1.2) {15} ;
\node[node_style] (diez) at (6.75,2.4) {10} ;
\node[node_style] (catorce) at (6.75,0) {14} ;

\draw (uno) -- (cinco);
\draw (uno) -- (nueve);
\draw (uno) -- (dos);
\draw (dos) -- (tres);
\draw (cuatro) -- (cinco);
\draw (cuatro) -- (seis);
\draw (cuatro) -- (siete);
\draw (cuatro) -- (ocho);
\draw (cuatro) -- (nueve);
\draw (cinco) -- (seis);
\draw (seis) -- (siete);
\draw (siete) -- (ocho);
\draw (ocho) -- (nueve);

\draw (quince) -- (diez);
\draw (quince) -- (once);
\draw (quince) -- (doce);
\draw (quince) -- (trece);
\draw (quince) -- (catorce);
\draw (diez) -- (once);
\draw (once) -- (doce);
\draw (doce) -- (trece);
\draw (trece) -- (catorce);
\draw (tres) -- (diez);
\draw (tres) -- (catorce);

\end{tikzpicture}
\end{center}

\caption{Example \ref{example_functionality_M}}
\end{figure}
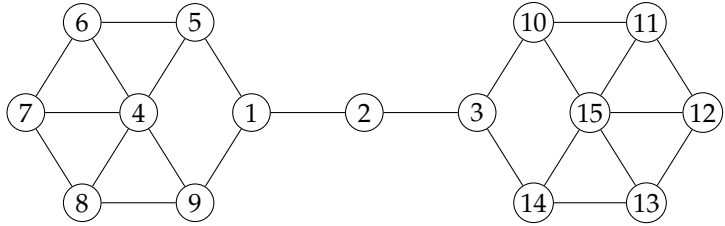

\begin{table}[h]
    \centering
    \begin{tabular}{c|c|c}
     \hline
     Nodes & Messages game & Attachment game
     \\
     \hline
        1,3 & 29.7 (14.16\%) & 2.5 (8.8\%) \\
        2 & 28.9 (13.72\%)  & 2.0 (7.1\%) \\
        4, 15 &  10.5 (5.01\%) & 2.1 (7.6\%) \\
        5,9,10,14 & 11.5 (5.49\%) & 1.8 (6.4\%) \\
        6,8,11,13 &  9.1 (4.34\%) &  1.6 (5.8\%) \\  
        7, 12 &  9.0 (4.28\%) & 1.6 (5.8\%) \\  
        \hline
    \end{tabular}
    \caption{Myerson centralities}
    \label{example_functionality_M}
\end{table}

When the purpose of the network is to send messages (see Table\ref{example_functionality_M}), node 2, which intermediates between the two subsocieties $S_1=\{1,4,5,6,7,8,9\}$ and $S_2=\{3,10,11,12,13,14,15\}$ is the second kind of node more central, whereas this second position is occupied by the hubs of $S_1$ and $S_2$, nodes 4 and 15 respectively, when the purpose is to get nodes to develop a joint project. Moreover, the differences between intermediary nodes 1, 2 and 3, and the remaining ones are considerably greater for the messages centrality.
\label{example-Myerson-functionality}
\end{example}

The Myerson value is characterized by means of two properties:

\begin{itemize}
    \item {\it Component efficiency:} For every game $(N,v)\in G^N$, and any social network $\Gamma=(N,E)\in {\cal G}^N$, it is verified:
    \begin{equation}
        \sum_{i\in S} \mu_i(N,v,E)=v(S),\; \text{ for all $S\in \K(\Gamma)$.}
        \label{component-efficiency}
    \end{equation}

    \item {\it Fairness:} For every game $(N,v)\in G^N$, and any social network $\Gamma=(N,E)\in {\cal G}^N$, it is verified:
    \begin{equation}
        \mu_i(N,v,E)-\mu_i(N,v,E\setminus \{e\})=\mu_j(N,v,E)-\mu_j(N,v,E\setminus \{e\}),\; \text{ for all $e=\{i,j\}\in E$.}
        \label{fairness}
    \end{equation} 
\end{itemize}

Component efficiency, which assures that all the value generated by a connected component is allocated to its nodes, allows for comparing centralities in different graph configurations that preserve the set of connected components (under the same functionality). Fairness, which is in fact the crucial property of the Myerson value, implies a symmetric effect of removing and edge over its both ends.

\section{Position centrality}

Similarly to the Myerson centrality, we propose an alternative family of centrality measures using the position value (Messens, 1988), in which the value assigned to a player depends on the value of the connections or links he has with other players. We will restrict to the subclass of zero-normalised symmetric and supperadditive games in order to obtain a minimum centrality equal to zero for every isolated node. 

Our main motivation is to escape from Myerson's property of fairness when the end nodes of the added edge have very different situations in the original social network in order to understand how the importance attributed to the links of a node affects its centrality measure. This is the case, for instance, of nodes 2 and 15 in Example \ref{example-Myerson-functionality}. It seems no reasonable that adding an edge between them affects both nodes in the same way.


First, we recall the definition of the position value restricted to this subclass of games in which the   family of position centrality measures is based. Then, we prove that the proposed family satisfies the most typical desirable properties of a centrality measure. In Section \ref{Adding} a thorough analysis of how the addition of an edge affects the position centrality of the nodes is made.


Let $CS_0^N$ denotes the class of all communication situations $(N, v, E)$ with player set $N$, being $(N, v)\in G^N_0$ a zero-normalised symmetric and supperadditive TU game, and $\G=(N,E)\in {\cal G}^N$ an undirected communication graph without loops or parallel edges. 
Let $(N,v,E)\in CS_0^N$, then the {\em link game} $(E,w^v)$ corresponding to $(N,v,E)$ (cf. Borm {\it et al.} 1992), which is played between edges instead of nodes, is defined as follows:
\begin{equation}
w^v(L)=\sum_{C\in \K(L) } v(C), \; \forall \, L\subseteq E ,
    \label{link-game-def}
\end{equation}
That is, the sum of the values of the maximal groups than can be formed when the only available connections are those of coalition $L$.

Then, the {\em Position value} $\pi: CS_0^N\longrightarrow \mathbb{R}^n$ is given by 
\begin{equation}
\pi_i(N,v,E)=\frac{1}{2} \sum_{e\in E_i} \phi_e(E,w^v), \; \forall \, i\in N,
    \label{position-value-def}
\end{equation}
where $\phi(E,w^v)$ denotes the Shapley value of the link game. 

The position value also admits an alternative expression in terms of the Harsanyi dividends of the link game (see Slikker, 2005), which is in fact very useful in order to prove many of the properties of the proposed family of centrality measures.

\begin{equation}
\pi _{i}\left( N,v,E\right) =\sum_{e\in E_{i}}\frac{1}{2}\sum_{\substack{L\subseteq E \\ e\in L}}\frac{\lambda_{L}\left(E, w^{v}\right) }{l}=\sum_{ L\subseteq E}\frac{1}{2}\lambda_{L}\left(E,w^{v}\right) \frac{l_{i}}{l},\text{ for all }i\in N,
\label{posicional-harsanyi-linkgame}
\end{equation}%
where $L_{i}=L\cap E_{i}$, and being $\lambda_{L}\left(E,w^{v}\right)$, the Harsanyi dividends of the link game $(E,w^{v})$. 

Slikker (2005) shows that $\lambda_{L}\left(E, w^{v}\right) =\lambda_L (w^v)$, that is, it does not depend on $E\supseteq L$. This facilitates the comparison of two communication situations that differ only in the underlying graph. Moreover, in the sequel, we will work with $\lambda_L(w^v)$ without any reference to the specific graph $(N,E)$.

\subsection{Properties of the Position centrality}
\label{propiedades_position_centrality}

In this context, we shall understand by {\em Position centrality} the position value of a given communication situation $(N,v,E)\in CS_0^N$. In the sequel $f(\cdot)$ will denote the function defining the symmetric game $(N,v)$. Now, we prove that the proposed family of Position centrality measures satisfies some of the most typical desirable properties of a centrality measure (see G\'omez {\it et al.}, 2003). In particular, it attains its maximum value for the hub of a star, which is according to Freeman (1979), a defining characteristic of proper centrality measures. Some proofs of these results are based on several lemmas  establishing interesting properties of the Harsanyi dividends of the link game, which are compiled in the Appendix. 

First, we prove that Position centrality is a non-negative centrality measure.  

\begin{proposition}
\label{positivity}
Let $(N,v,E)\in CS_0^N$, then 
$\pi_{i}(N,v,E)\geq 0$, for all $i\in N$.
\end{proposition}

\begin{proof}
The proof strongly depends on the superadditivity of the game. By definition 
\begin{equation*}
\pi _{i}\left( N,v,E\right) =\frac{1}{2}\sum_{e\in
E_{i}}\phi _{e}(E,w^v),
\end{equation*}%
where 
\begin{equation*}
\phi _{e}(E,w^v)=\sum_{L\subseteq E\backslash \{e\}}\frac{l!( \vert E\vert -l-1)!}{\vert E\vert !}\left( w^{v}(
L\cup \{e\}) -w^{v}( L)\right ), \text{ with $ e=\{i,j\}$.} 
\end{equation*}%
Let $\mathcal{K}(L)$ be
the set of connected components. W.l.o.g. suppose that $\mathcal{K}(L)$ has
only two elements, i.e., $\mathcal{K}(L)=\left\{ T_{1},T_{2}\right\} $ and $i\in T_{1} $. To
examine the sign of $w^{v}\left( L\cup \{e\}\right)
-w^{v}\left( L\right) $, we need to consider the
following three cases:
\begin{enumerate}
\item $j\in T_{1}$. Then clearly $w^{v}\left( L\cup
\{e\}\right) -w^{v}\left( L\right) =0.$\newline
\item $j\in T_{2}$. Then we have:%
\begin{equation*}
w^{v}\left( L\cup \{e\}\right) -w^{v}\left( L\right) =v\left( N[T_{1}\cup T_{2}]\right) -v\left( N[T_{1}]\right) -v\left(
N[T_{2}]\right) \geq 0\text{.}
\end{equation*}
by superadditivity of $v.$\newline
\item $j\notin  T_{1}\cup  T_{2}$ . Then:%
\begin{equation*}%
w^{v}\left( L\cup \{e\}\right) -w^{v}\left( L\right) =w^{v}(E[T_{1}]\cup  \{e\})-w^{v}(E[T_{1}]) =v(T_1\cup \{j\})-v(T_1)\geq 0,
\end{equation*}
since $(N,v)$ is supperaditive and zero-normalised.
\end{enumerate}
\end{proof}

Next, we show that it verifies the property of {\em Locality}, which establishes that the centrality of a node depends only on the connected component to which it belongs. Moreover, it also satisfies {\em Component efficiency}. The following lemma, which generalises the analogous result of Borm {\em et al.} (1992)  for cycle-free graphs, is needed.

\begin{lemma}
Let $\left( N,v\right) $ be a $TU$ game and $\Gamma =\left( N,E\right) \in 
\mathcal{G}^{N}$. For every subset $L\subseteq E$ such that $\Gamma_L$ is not connected it holds $\lambda_{L}\left(w^{v}\right) = 0$.
\label{dividendos-0-noconexo}
\end{lemma}

\begin{proposition}
Position centrality verifies Locality, i.e. for every communication situation $(N,v,E)\in CS_0^N$, and every node $i\in N$ it holds:
$$
\pi_{i}(N,v,E)=\pi_{i}(\K_i(E),v_i,E[\K_i(E)]),
$$
where $v_i$ is the restriction of $v$ to $\K_i(E)$.
\label{locality}
\end{proposition}

\begin{proof}
Let $\mathcal{K}(E)$ be the set of connected components in the graph $(N,E)$
and\thinspace $i\in N$. W.l.o.g. suppose that $\mathcal{K}(E)$ has only two
elements, i.e., $\mathcal{K}(E)=\left\{ \mathcal{K}_{i}(E),T\right\} $. We are going to calculate  $\pi _{i}\left( N,v,E\right) $ using (\ref{posicional-harsanyi-linkgame}):
\begin{equation*}
\pi _{i}\left( N,v,E\right)  =\sum_{L\subseteq E}\frac{1}{2}\lambda
_{L}\left( w^{v}\right) \frac{l_{i}}{l} .
\end{equation*}
By Lemma \ref{dividendos-0-noconexo}, last expression can be decomposed as follows: 
\begin{equation*}
\pi _{i}\left( N,v,E\right) =\sum_{L\subseteq E\left[ \mathcal{K}_{i}(E)\right] }\frac{1}{2}\lambda
_{L}\left( w^{v}\right) \frac{l_{i}}{l}+\sum_{L\subseteq E\left[ T\right] } \frac{1}{2}\lambda _{L}\left( w^{v}\right) \frac{l_{i}}{l} .
\end{equation*}
Since $L_{i}\subseteq $ $E\left[ \mathcal{K}_{i}(E)\right]$, it follows that
$L_{i} \cap E\left[ T\right] = \varnothing $. Thus,

\begin{equation*}
\pi _{i}\left( N,v,E\right) =\sum_{L\subseteq E\left[ \mathcal{K}_{i}(E)%
\right] }\frac{1}{2}\lambda _{L}\left( w^{v}\right) \frac{l_{i}}{l}  =\sum_{L\subseteq E\left[ \mathcal{K}%
_{i}(E)\right] }\frac{1}{2}\lambda _{L}\left( w^{v_{i}}\right) \frac{l_{i}}{l%
},
\end{equation*}%
which completes the proof. 
\end{proof}

\begin{proposition}
\label{ComponentEfficiency}
Position centrality verifies Component efficiency.
\end{proposition}

\begin{proof}
Component efficiency is proven by Borm et al. (1992).
\end{proof}

Note that Component efficiency, restricted to the class $CS_0^N$, allows for comparing centralities in different graph configurations that preserve the number of connected components and their cardinalities (under the same functionality $f$).

The following propositions show results about the minimal and the maximal position centrality of a node. Isolated nodes have minimal position centrality, which equals 0, whereas maximal centrality is attained by the hub of a star.

\begin{proposition}
\label{isolatedmin}
Let $i_0\in N$ be an isolated node in the graph $(N,E)$, then 
$$
\pi_{i_0}(N,v,E)=0\leq \pi_i(N,v',E'),
$$
for all $i\in N$, and every  $(N,v,E), (N,v',E')\in CS_0^N$.
\end{proposition}

\begin{proof}
Let $i_0\in N$ be an isolated node in $(N,E)$, then it follows from (\ref{posicional-harsanyi-linkgame}) and the fact that $l_{i_{0}}=0$ for all $L\subseteq E$, that: 
\begin{equation*}
\pi _{i_{0}}\left( N,v,E\right) =\sum_{L\subseteq E}\frac{1}{2}\lambda
_{L}\left( w^{v}\right) \frac{l_{i_{0}}}{l}=0, \text{ for all $(N,v)\in G_0^N$,}
\end{equation*}%
which is the minimum position centrality of a node, taking into account Proposition \ref{positivity}.
\end{proof}
 
\begin{proposition}
\label{starmax}
Let $(N,v)\in G^N_0$. If $(N,E^{*})\in \mathcal{G}^{N}$ is the star with $n$ nodes where node $1$ is the hub, then for all connected graph $(N,E)\in \mathcal{G}^{N}$, and for all $i\in N$ is verified:
\begin{equation*}
\pi_{1}(N,v,E^{*})\geq \pi _{i}(N,v,E).
\end{equation*}
\end{proposition}

\begin{proof}
 Let $(N,E^{*})\in \mathcal{G}^{N}$ be the star with $n$ nodes. By symmetry of the Shapley value we have that $\phi _{e}(E^{*},w^{v})=\frac{%
f(n)}{n-1}$ for any $e$ in $E^{*}$ and therefore $\pi _{1}\left( N,v,E^{*}\right) =\frac{f(n)}{2}$.
Now suppose that there exists a graph $(N,E)$ and a node $i\in N$, such that 
\begin{equation*}
\pi _{1}\left( N,v,E^{*}\right)<\pi _{i}\left( N,v,E\right)\text{,}
\end{equation*}%
then 
\begin{equation*}
\pi _{i}\left( N,v,E\right) > \frac{f\left( n\right) }{2}\text{.}
\end{equation*}%
On the other hand, by definition%
\begin{equation*}
\pi _{i}\left( N,v,E\right)=\frac{1}{2}\sum_{e\in E_i} \phi_e(E,w^v), \; \forall \, i\in N
\end{equation*}%
and consequently 
\begin{equation*}
      \sum_{e\in E} \phi_e(E,w^v)>f\left( n\right)  \text{,}
\end{equation*}
which contradicts the efficiency property of Shapley value.
\end{proof}

We will now examine in Propositions \ref{min_chain}, \ref{order_chain} and \ref{min_connected} some properties related to a chain. The Position centrality of an end node of a chain increases with the length of the chain. For a given chain, the Position centrality of its nodes is maximal for the middle nodes and decreases symmetrically from the middle to the end nodes. Moreover, proposition \ref{isolatedmin} is reinforced,  the minimal Position centrality in connected graphs is attained by the end nodes of a chain.

First, some general results about the Harsanyi dividends of the link game when the underlying graph contains no cycles are given.


\begin{lemma}
Let $(N,v)\in G^N_0$ and $\Gamma =\left( N,E\right) \in \mathcal{G}^{N}$. If $\Gamma$  is cycle-free, then for all $\emptyset \neq L\subseteq E$ with $\Gamma_L$ connected, the Harsanyi dividends of the link game can
be calculated in the following way:%
\begin{equation}
\lambda _{L}\left( w^{v}\right) =F\left( l+1,l-d_{L}\right),
\label{funcionFdividendos}
\end{equation}%
where $d_{L}=\left\vert \mathcal{D}\left( L\right)\right\vert $, being $F(s,r)= \sum_{k=0}^{r}\left( -1\right) ^{k}\binom{r}{k}f\left(s-k\right)$, for $s, r\in \mathbb{N}$ and $s\geq r$.
\label{dividendos-exp-F}
\end{lemma}

\begin{figure}[h]
    \begin{center}
    \includegraphics{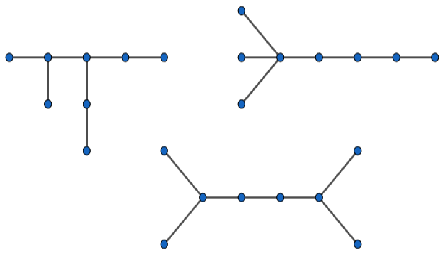}
    \end{center}
    \caption{Graphs with the same Harsanyi dividend}
   \label{hola}
    \end{figure}

Note that expression \eqref{funcionFdividendos} only depends on $l$,  $d_L$ and $f$. Thus, all subset of edges that form a cycle-free graphs with equal number of edges and extreme points (number of edges minus number of cutedges) have the same dividend for any fixed game. For example, the three graphs in Figure \ref{hola}  have the same Harsanyi dividend since they 
have 7 edges  and 4 extreme points. Then, in all cases we have:

\begin{equation*}
\lambda _{L}\left( w^{v}\right) =F(8,4)=f\left( 8\right) -4f\left( 7\right)
+6f\left( 6\right) -4f\left( 5\right) +f\left( 4\right) \text{.}
\end{equation*}

Last lemma will allow us to simplify the calculation of dividends for chains, compare the dividends of trees with $l$ edges but with different structures, and determine the sign of the dividend under additional conditions for the game.

\begin{lemma}
Let $f$ be a real function such that $f^{\left(
k\right) }\left( x\right) \geq 0$ in $[1,+\infty)$, for  $k=0,1,...,n$, then  the function $F(s,r)$ is decreasing in the second argument for every fixed $s\in \mathbb{N}$, with $r\in \mathbb{N}, s\geq r$.
\label{coro-comparando-div}
\end{lemma}

Next corollary follows straightforwardly from Lemmas \ref{dividendos-exp-F} and \ref{coro-comparando-div}. 

\begin{corollary}
Let $(N,v)\in G^N_0$ and $\Gamma =\left( N,E\right) \in \mathcal{G}^{N}$. If $\Gamma$  is cycle-free, $f\in \mathcal{C}^{n}\left( \mathbb{R}\right) $ and $f^{\left(
k\right) }\left( x\right) \geq 0$ in $[1,+\infty)$, for  $k=0,1,...,n$, then, for every pair of subsets $L,L^{\prime }\subseteq E$, such that $l=l^{\prime }$ and $d_{L}\leq d_{L^{\prime }}$ it holds:
$$
 \lambda_{L}(w^{v})\leq \lambda_{L'}(w^{v})
 $$
 \label{coro-comp-dividendos-cutedges}
\end{corollary}

In particular, if $f$ satisfies the required conditions of Corollary \ref{coro-comparando-div}, the star is the structure with the lowest Harsanyi dividend, whereas the chain is the one with the highest value, among all the $n$-node trees. 
That is:
\begin{equation*}
\lambda_{L^{*}}\left( w^{v}\right) \leq \lambda%
_{L}\left( w^{v}\right) \leq \lambda_{L^{C}}\left( w^{v}\right),
\end{equation*}%
where $L^{*}$ and $L^{C}$ are two sets of edges that form a star and a chain, respectively,  and being $L$ a set of edges forming a tree with $\left\vert L^*\right\vert =\left\vert L\right\vert=\left\vert L^{C}\right\vert $.

With respect to the sign of the dividends, if the graph $\Gamma$ is a chain with $l$ edges, taking into account \eqref{funcionFdividendos}, the Harsanyi dividends are given by:
\begin{equation*}
\lambda_{L}\left( w^{v}\right) =F(l+1,2)=f\left( l+1\right) -2f\left(
l\right) +f\left( l-1\right),
\end{equation*}
which are always non negative if the game $(N,v)$ is convex. However, for general cycle-free graphs, convexity of $f$ must be strengthened in order to assure non negative dividends.

\begin{corollary}
Let $(N,v)\in G^N_0$ and $\Gamma =\left( N,E\right) \in \mathcal{G}^{N}$. If $\Gamma$  is cycle-free, $f\in \mathcal{C}^{n}\left( \mathbb{R}\right) $ and $f^{\left(
k\right) }\left( x\right) \geq 0$ in $[1,+\infty)$, for  $k=0,1,\dots,n$, then $\lambda _{L}\left( w^{v}\right) \geq 0$, for all $\emptyset \neq L\subseteq E$.
\label{dividendo-positivo-fcond}
\end{corollary}

\begin{proposition}
Let $(N,v)\in G^N_0$ be convex. Let us suppose that $(N_k,E_{k}^{C})$ is a chain with $n_{k}$
nodes ordered in the natural way. If $n_{1}<n_{2}$, then:%
\begin{equation*}
\pi_{1}\left(N_1,v,E_{1}^{C}\right) \leq \pi_{1}\left( N_2,v,E_{2}^{C}\right) 
\text{.}
\end{equation*}
\label{min_chain}
\end{proposition}

\begin{proof}
We will show that the result is true for $n_{2}=n_{1}+1$. Applying \eqref{posicional-harsanyi-linkgame}, we have that:%
\begin{equation*}
\pi_{1}\left( N_{k},v,E_{k}^{C}\right) =\frac{1}{2}\sum_{L\subseteq E_{k}^{C}}%
\frac{\lambda_{L}\left( w^{v}\right) }{l}=\frac{1}{2}\sum_{l < n_{k}}\frac{F(l+1,2)}{l}\text{,}
\end{equation*}%
where the second equality follows from Lemma \ref{dividendos-0-noconexo}, since only the subset of edges $L$ that form a chain have a non-zero  Harsanyi dividend, and Lemma \ref{dividendos-exp-F}. %

Consequently
\begin{equation*}
\pi_{1}\left( N_{2},v,E_{2}^{C}\right) -\pi_{1}\left( N_{1},v,E_{1}^{C}\right) =%
\frac{1}{2}\left( \sum_{l < n_{2}}\frac{F(l+1,2)}{l}-\sum_{l < n_{1}}%
\frac{F(l+1,2)}{l}\right) =\frac{1}{2}\frac{\lambda _{E_{2}^{C}}\left(
w^{v}\right) }{n_{2}-1}\text{.}
\end{equation*}%
Finally, the result follows since $\lambda _{E_{2}^{C}}\left( w^{v}\right)
\geq 0$ as a consequence of the convexity of $v$.
\end{proof}

\begin{proposition}
If $(N,v)\in G^N_0$ is convex and $(N,E^{C})$ is a chain with $n$ nodes numbered in the natural way, then for $1\leq i\leq n/2:$%
\begin{equation*}
\pi_{i}\left(N,v,E^{C}\right) \leq \pi _{i+1}\left(N,v,E^{C}\right) .
\end{equation*}
\label{order_chain}
\end{proposition}

\begin{proof}
Since $E^C$ remains fixed along the proof we will write $\phi_e(E^C,w^V)$ simply as $\phi_e(w^V)$. 

Let $e_i=\{i,i+1\}$, $i=1,\dots,n-1$. By definition:%
\begin{eqnarray*}
\pi _{i}\left( N,v,E^{C}\right)  &=&\frac{1}{2}\left( \phi_{e_{i-1}}(w^{v})+\phi _{e_{i}}(w^{v})\right) \text{ and} \\
\pi_{i+1}\left( N,v,E^{C}\right)  &=&\frac{1}{2}\left( \phi_{e_{i}}(w^{v})+\phi_{e_{i+1}}(w^{v})\right) \text{.}
\end{eqnarray*}%
%
Now, we prove a recursive formula for the Shapley values of the
edges: 
\begin{equation}
\phi_{e_{j}}(w^{v})=\phi _{e_{j-1}}(w^{v})+\sum_{k=j}^{n-j}%
\frac{F\left( k+1,2\right) }{k}, \text{ for }1\leq j\leq n/2\text{,}
\label{exp-recursiva-cadenas}
\end{equation}
assuming as initial condition $\phi_{e_{0}}(w^{v}):=0$. 

For $j=1$, $\phi_{e_{1}}(w^{v})$ can be derived from expression \eqref{posicional-harsanyi-linkgame}, taking into account that only the subsets $L$ such that $\Gamma _{L\text{ }}$ is a chain have a non-zero dividend. Thus,
\begin{equation*}
\phi_{e_{1}}(w^{v})=\sum_{e_{1}\in L\subseteq E^C}\frac{\lambda_{L}\left( w^{v}\right) }{l}=\sum_{k=1}^{n-1}\frac{F\left( k+1,2\right) }{k}%
\text{.}
\end{equation*}
For $j\geq 2$, note that there is one coalition of size $1$ containing $e_{j}$, two coalitions of size $2$, three of size $3$, and so on increasing successively up to size $j$; from this point, for all $s\leq n-j$ there are always exactly $j$ coalitions of size $s$. Finally, for all $n-j<s\leq n-1$, the number of coalitions of size $s$ decreases by $1$ starting at $n-j+1$. Therefore:
\begin{equation*}
\phi_{e_{j}}(w^{v})=\sum_{k=1}^{j}k\frac{F\left( k+1,2\right) }{k}+j\sum_{k=j+1}^{n-j}\frac{F\left( k+1,2\right) }{k}+\sum_{k=n-j+1}^{n-1}(n-k)\frac{F\left( k+1,2\right) }{k}\text{.}
\end{equation*}
Then, it can be  obtained directly that 
$$
\phi_{e_{j}}(w^{v}) - \phi_{e_{j-1}}(w^{v}) = \sum_{k=j}^{n-j}\frac{F\left( k+1,2\right) }{k}\geq 0,
$$
since $(N,v)$ is convex. Thus, 
$$
\pi_{i+1}\left( N,v,E^{C}\right) -\pi_{i}\left( N,v,E^{C}\right) =
\frac{1}{2}(\phi_{e_{i+1}}(w^{v})- \phi_{e_{i}}(w^{v})+ \phi_{e_{i}}(w^{v}) -\phi_{e_{i-1}}(w^{v})) \geq 0 .
$$
\end{proof}

\begin{proposition}
If $(N,v)\in G^N_0$ is convex and $(N,E^{C})$ is the chain with $n$ nodes,
where node $1$ is an end node, then for all connected graphs $(N,E)\in \mathcal{G}_0^N$, and for all $i\in N$, it holds:%
\begin{equation*}
\pi_{1}(N,v,E^{C})\leq \pi_{i}(N,v,E)
\end{equation*}
\label{min_connected}
\end{proposition}

\vspace*{-10mm}
\begin{proof}
The proof follows the same lines as the proof of Proposition 3.5 in G\'omez {\it et al.} (2003).
\end{proof}

\section{Adding an edge}
\label{Adding}

In the analysis of the family of centrality measures derived form the use of the Myerson value (G\'omez {\it et al.}, 2003) the symmetrical behaviour with respect to the addition or elimination of an edge is a fundamental characteristic. However, the position centrality measure, unlike the Myerson centrality measure, responds in a more versatile way to such addition or elimination.

Next, we show some situations in which adding an edge benefits both end nodes, although not necessarily to the same amount. However, examples in which one of the end nodes becomes worse can be found. 


\begin{proposition} Let $(N,v)\in G^N_0$ be a convex game, and let $(N_1,E_1)$ and $(N_2,E_2)$ be two not connected graphs, with $N=N_1\cup N_2$. If a bridge $b=\{i_0,j_0\}$, with $i_0\in N_1$ and $j_0\in N_2$ is added, then it holds:
\begin{equation}
\pi_i(N_k,v_k,E_k)=\pi_i(N,v,E)\leq \pi_i(N,v,E\cup\{ b\}), 
\label{adding-bridge}
\end{equation}
for all node $i\in N_k$, $k=1,2$, where $v_k$ denotes the restriction of $v$ to $N_k$, $k=1,2$, and $E=E_1\cup E_2$.
\label{adding-bridge-prop}
\end{proposition}


\begin{proof}
To shorten notation, we write $w$ instead of $(E,w^v)$, and $w^{+}$ instead of $(E\cup \{b\},w^v)$. 

Note that the first equality of \eqref{adding-bridge} holds since the position value verifies {\em Locality} (Proposition \ref{locality}).  We will prove the second inequality, i.e., $\pi_i(N,v,E)\leq \pi_i(N,v,E\cup\{ b\})$, for all $i\in N$, by verifying that $\phi_e (w)\leq \phi_{e} (w^{+})$ for all $e \in E$. 

Let $e \in E$ be an original edge, then it holds 
\begin{equation}
  \phi_e (w):= \sum_{L\subseteq E\setminus \{e\}} \frac{l! (m-l-1)!}{m!} \bigl ( w(L\cup \{e\})-w(L)\bigr ),   
    \label{phi_ell_sin}
\end{equation}
and
\begin{multline}
  \phi_e (w^{+})= \sum_{L\subseteq E\setminus \{e\}} \Bigl \{ \frac{l! (m-l)!}{(m+1)!} \bigl ( w^{+}(L\cup \{e\})-w^{+}(L)\bigr ) +
  \\[2mm]
    \frac{(l+1)! (m-l-1)!}{(m+1)!} \bigl ( w^{+}(L\cup \{e,b\})-w^{+}(L\cup \{b\})\bigr ) \Bigr \},
    \label{phi_ell_con}
\end{multline}
where $m=\abs{E}$.

Let $e\in E_1$ (the same argument applies to $e \in E_2$) and $L\subseteq E$. Clearly, 
\begin{equation}
    w^{+}(L\cup \{e\})-w^{+}(L)  =  w(L\cup \{e\})-w(L), \forall\, L\subseteq E . 
    \label{phi_ell_desigualdad_bridge_sin}   
\end{equation}
%
Now, we will prove that
\begin{equation}
    w^{+}(L\cup \{e,b\})-w^{+}(L\cup \{b\})  \geq  w(L\cup \{e\})-w(L) .
\label{phi_ell_desigualdad_bridge_con}    
\end{equation} 
We will distinguish two cases: 
\begin{enumerate}[$(i)$]
\item If $i_0\notin \K_i (L\cup\{e\})=\K_j((L\cup\{e\})$, with $e=\{i,j\}$ (note that then $e\notin E_{i_0}$), then $ \K_i (L\cup\{e\}) =  \K_i (L\cup\{e,b\})$ and thus $w^{+}(L\cup \{e,b\})-w^{+}(L\cup \{b\})  =  w(L\cup \{e\})-w(L)$.
\item If $i_0\in \K_i (L\cup\{e\})$, then $\K (L\cup\{ e,b\})$ equals $\K (L\cup\{ e\})$ replacing 
 $\K_{i_0} (L \cup\{e\})=\K_i (L\cup\{e\})$ by $\K_i (L\cup\{e\})\cup \R_{j_0}(L)$, where
$$
\R_{j_0}(L)=\begin{cases}  \{ j_0\}, & \text{ if $L\cap E_{j_0}=\varnothing$,}
\\
\K_{j_0} (L)=\K_{j_0} (L\cup \{ e\}), & \text{ otherwise.}
\end{cases}
$$
Note that the former component $\K_{j_0} (L)$ is replaced by $\K_{i_0} (L \cup\{e\})$ if $L\cap E_{j_0}\neq \varnothing$. Analogously, $\K (L\cup\{ b\})$ equals $\K (L)$ replacing 
 $\K_{i_0} (L)$ by $\K_{i_0} (L)\cup \R_{j_0}(L)$.
 Thus, 
 \begin{multline*}
   \Bigl ( w^{+}(L\cup \{e,b\})-w^{+}(L\cup \{b\}) \Bigr ) -\Bigl (  w(L\cup \{e\})-w(L) \Bigr ) =
  \\
  \Bigl ( v\bigl (\K_i (L\cup\{e\})\cup \R_{j_0}(L)\bigr )- v\bigl(\K_i (L)\cup \R_{j_0}(L)\bigl)\Bigr )  - \Bigl ( v\bigl (\K_i (L\cup\{e\})\bigr )-v \bigl(\K_i (L) \bigr )
  \Bigr ) ,
 \end{multline*}
which is non-negative since $(N,v)$ is convex, 
and thus \eqref{phi_ell_desigualdad_bridge_con}  is verified. 
\end{enumerate}

Therefore, taking into account \eqref{phi_ell_desigualdad_bridge_sin} and \eqref{phi_ell_desigualdad_bridge_con} into expression \eqref{phi_ell_con} it follows that $\phi_e (w^{+})\geq \phi_e (w)$, for all $e\in E$. Thus, for all $i\neq i_0,j_0$
$$
\pi_i(N,v,E\cup\{ b\}):=\frac{1}{2} \sum_{e \in E_i} \phi_e (w^{+}) \geq \frac{1}{2} \sum_{e \in E_i} \phi_e (w):= \pi_i(N,v,E) .
$$
For the end nodes $i_0,j_0$ of the bridge, there is a common contribution given by $\frac{1}{2} \phi_{b} (w^{+}) \geq 0$ (see Proposition \ref{positivity} proof), and -contrary to the Myerson value- an independent increasing that depends on 
how the network structure newly added to each original independent component affects to the original connections of $i_0$ and $j_0$. To be specific,
\begin{align*}
\pi_{i_0}(N,v,E\cup\{ b\})-\pi_{i_0}(N,v,E) & = \frac{1}{2} \phi_{b} (w^{+})  +
 \frac{1}{2} \sum_{e \in E_{i_0}} (\phi_e(w^{+}) - \phi_e(w)),
 \\
\pi_{j_0}(N,v,E\cup\{ b\})-\pi_{j_0}(N,v,E) & = \frac{1}{2} \phi_{b} (w^{+})  +
 \frac{1}{2} \sum_{e \in E_{j_0}} (\phi_e(w^{+}) - \phi_e(w)) .
\end{align*}
\end{proof}

\begin{example}
Adding the bridge $b=\{1,4\}$ to the not connected social network depicted in Figure \ref{puente-2desconectadas-SN} increases the centrality of every node. However, the gain of each node varies, if for instance the messages game is considered, node 4 gains  more position centrality than node 1 ($6.5$ versus $5.6$), which is even more pronounced in relative terms (433.3\% versus 280\%).
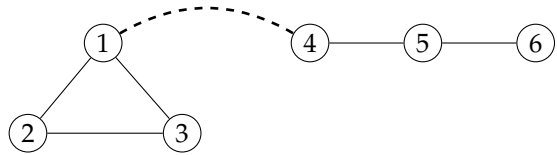
\begin{figure}[h]
\begin{center}
\begin{tikzpicture}[node_style/.style={draw,circle,minimum size=0.5cm,inner sep=1}]
\node[node_style] (uno) at (9,1.2) {1} ;
\node[node_style] (dos) at (8.0,0.0) {2} ;
\node[node_style] (tres) at (10.05,0.0) {3} ;
\node[node_style] (cuatro) at (11.75,1.2) {4} ;
\node[node_style] (cinco) at (13.25,1.2) {5} ;
\node[node_style] (seis) at (14.75,1.2) {6} ;
\draw (uno) -- (dos);
\draw (dos) -- (tres);
\draw (tres) -- (uno);
\draw (cuatro) -- (cinco);
\draw (cinco) -- (seis);
\draw [dashed,line width=1.0pt] (uno) edge[-,bend left=30]  (cuatro);
\end{tikzpicture}
\end{center}
\caption{Not connected societies example.}
\label{puente-2desconectadas-SN}
\end{figure}

\label{puente-2desconectadas-ejemplo}
\end{example}

For the specific case in which the two not connected graphs are two stars, and asking for some extra property of the functionality, the previous result can be strengthened. 

\begin{proposition}
Let $(N,v)\in G^N_0$ defined by a function $f(\cdot)\in {\cal C}^n(\mathbb{R})$ such that $f^{(k)}(x)\geq 0$ in $[1,+\infty)$, for all $k=1,\dots,n$.
Let $(N_1,E_1^{*})$ and $(N_2,E_2^{*})$ be two not connected stars of $k_1$ and $k_2$ leaves respectively, with $k_1\leq k_2$, and being $N=N_1\cup N_2$ and $E=E_1^{*}\cup E_2^{*}$. If a bridge $b=\{c_1,c_2\}$  between the two hubs $c_1\in N_1$ and $c_2\in N_2$ is added, then it holds that:
\begin{enumerate}
     \item $\pi_{c_1}(N,v,E\cup\{ b\})-\pi_{c_1}(N,v,E) \leq \pi_{c_2}(N,v,E\cup\{ b\})-\pi_{c_2}(N,v,E)$, 
     \item $\pi_{i}(N,v,E\cup\{ b\})-\pi_{i}(N,v,E) \geq \pi_{j}(N,v,E\cup\{ b\})-\pi_{j}(N,v,E)$, for every pair of leaves $i\in N_1\setminus{c_1}$, $j\in N_2\setminus{c_2}$. 
\end{enumerate}

\end{proposition}

\begin{proof}
Let us first prove that the hub of the larger star gains more position value than the hub of the smaller one. 

From expression \eqref{posicional-harsanyi-linkgame} of the Position value in terms of the Harsanyi dividends of the link game proposed in Slikker (2005), it follows that:
\begin{align}
\Delta \pi _{c_{1}} &=\pi _{c_{1}}(N,v,E\cup \{b\})-\pi _{c_{1}}(N,v,E)=\frac{1}{2}\sum_{L\subseteq E}
\lambda_{L\cup \{b\}}\frac{l_{c_{1}}+1}{l+1}\text{,}
\label{inc-hub-peq}
\\
\Delta \pi _{c_{2}} & =\pi _{c_{2}}(N,v,E\cup \{b\})-\pi _{c_{2}}(N,v,E)=\frac{%
1}{2}\sum_{L\subseteq E}\lambda _{L\cup \{b\}}\frac{l_{c_{2}}+1}{l+1},
\label{inc-hub-big}
\end{align}%
where, throughout the proof, $\lambda_L$ denote the Harsanyi dividends of the link game $w^v$. Note that the  addends with non-zero dividend in the above two sums correspond to coalitions $L$ of the form:
\begin{itemize}
    \item[(s0)] $L=\{b\}$,
    \item[(s1)]  $L=T_1\cup \{b\}$, with $\emptyset\neq T_1\subseteq E_1^{*}$,
    \item[(s2)]  $L=T_2\cup \{b\}$, with $\emptyset\neq T_2\subseteq E_2^{*}$,
    \item[(s3)]  $L=T_1\cup T_2\cup \{b\}$, with $\emptyset \neq  T_1\subseteq E_1^{*}$ and $\emptyset\neq T_2\subseteq E_2^{*}$.
\end{itemize}
In the unique type (s0) addend, the dividend $\lambda_{\{b\}}$ appears in both increments \eqref{inc-hub-peq} and \eqref{inc-hub-big} with the same coefficient 1, whereas for the remaining types, the dividends appear with different coefficients. However, taking into account  Lemma \ref{dividendos-exp-F}, $\lambda_{T_1\cup \{b\}}=\lambda_{T_2\cup \{b\}}$, whenever  $\abs{T_1}=t_1=t_2=\abs{T_2}$, all type (s1) addends have an equal one of type (s2) with $T_2$ s.t. $t_2\leq k_1$. Analogously, $\lambda_{T_1\cup T_2\cup \{b\}}=\lambda_{T'_1\cup T'_2\cup \{b\}}$, if $t_1+t_2=t'_1+t'_2$, and therefore, all type (s3) addends for $T_2$ with $t_2\leq k_1$ appear in both increments. Thus,  their difference $ \Delta \pi_{c_{2}}-\Delta \pi_{c_{1}}$ is given by:
%
$$
 \frac{1}{2}\sum_{\substack{\emptyset\neq T_2\subseteq E_2^{*} \\
    t_2>k_1}} \lambda_{T_2\cup \{b\}} \Bigl ( 1-  \frac{1}{t_2+1}\Bigr ) +  \frac{1}{2} \sum_{\substack{\emptyset\neq T_1\subseteq E_1^{*} \\
    T_2\subseteq E_2^{*}, t_2>k_1}} \lambda_{T_1\cup T_2\cup \{b\}} \Bigl ( \frac{t_2+1}{t_1+t_2+1}-  \frac{t_1+1}{t_1+t_2+1}\Bigr ) ,
$$
which is non-negative by Corollary \ref{dividendo-positivo-fcond}, and taking into account that $t_2>k_1\geq t_1$.

With respect to the second statement, let $i\in N_{1}\backslash c_{1}$ and $j\in N_{2}\backslash c_{2}$, be any pair of leaves. Analogously, we must consider the following sort of non-zero addends in their respective increments:
\begin{itemize}
    \item[(s1)] $L=T_1\cup \{\{i,c_1\},b\}$,$T_1\subseteq E_1^{*}\setminus \{i,c_1\}$,
    \item[(s2)]  $L=T_2\cup \{\{j,c_2\},b\}$,$T_2\subseteq E_2^{*}\setminus \{j,c_2\}$,
    \item[(s3$i$)]  $L=T_1\cup T_2\cup \{\{i,c_1\}, b\}$, $T_1\subseteq E_1^{*}\setminus \{i,c_1\}$ and  $\emptyset  \neq T_2\subseteq E_2^{*}\setminus \{j,c_2\}$,
    \item[(s3$j$)]  $L=T_1\cup T_2\cup \{\{j,c_2\}, b\}$, $\emptyset  \neq T_1\subseteq E_1^{*}\setminus \{i,c_1\}$ and  $T_2\subseteq E_2^{*}\setminus \{j,c_2\}$,
    \item[(s4)]  $L=T_1\cup T_2\cup \{\{i,c_1\},\{j,c_2\},b\}$, with $T_1\subseteq E_1^{*}\setminus \{i,c_1\}$ and $T_2\subseteq E_2^{*}\setminus \{j,c_2\}$.
\end{itemize}
In this case, comparing both increments is more subtle because it requires to compare the dividends of two different structures:


\begin{multicols}{2}
\begin{center}
\begin{tikzpicture}[node_style/.style={draw,circle,minimum size=0.5cm,inner sep=1}]
\node[node_style] (ci) at (1.5,1.2) {$c_1$} ;
\node[node_style] (uno) at (4.5,1.2) {1} ;
\node[node_style] (j) at (4.2,2.2) {$j$} ;
\node[node_style] (cj) at (3.25,1.2) {$c_2$} ;
\node[node_style] (dos) at (3.00,2.4) {2} ;
\draw [dashed,line width=1.0pt] (ci) -- (cj);
\draw (cj) -- (j);
\draw (cj) -- (uno);
\draw (cj) -- (dos);
\end{tikzpicture}
\\[4mm]
Type (s2) $L$, with $l=4, d_L=0, \lambda_L=F(5,4)$
\end{center}

\begin{center}
\begin{tikzpicture}[node_style/.style={draw,circle,minimum size=0.5cm,inner sep=1}]
\node[node_style] (i) at (1.5,2.4) {$i$} ;
\node[node_style] (ci) at (1.5,1.2) {$c_1$} ;
\node[node_style] (uno) at (4.5,1.2) {1} ;
\node[node_style] (j) at (4.2,2.2) {$j$} ;
\node[node_style] (cj) at (3.25,1.2) {$c_2$} ;
\draw (ci) -- (i);
\draw [dashed,line width=1.0pt] (ci) -- (cj);
\draw (cj) -- (j);
\draw (cj) -- (uno);
\end{tikzpicture}
\\[4mm]
Type (s3$i$) $L$, with $l=4, d_L=1, \lambda_L=F(5,3)$
\end{center}

\end{multicols}

\vspace*{5mm}
 Type (s1) addends appear only in $\Delta\pi_i$, whereas  type (s2)  addends appear only in $\Delta\pi_j$. Again, by Lemma \ref{dividendos-exp-F}, every type (s1) addend in 
$\Delta\pi_i$ has an equal one of type (s2) with $T_2$ s.t. $t_2\leq k_1$ in $\Delta\pi_j$. With respect to type (s3$i$) and (s3$j$) addends, both of them appear in both increments with the exception of the extreme cases: $T_1=\emptyset$ in type (s3$i$) addends and $T_2=\emptyset$ in type (s3$j$) addends, which again also coincide whenever $t_2\leq k_1-1$ in the extreme case $T_1=\emptyset$. Type (s4) addends appear in both cases. Therefore,  it follows from expression \eqref{funcionFdividendos} that:
\begin{align*}
    \Delta \pi_i -\Delta \pi_j = & \frac{1}{2} \sum_{\substack{T_2\subseteq E_2^{*}\setminus \{j,c_2\} \\
    t_2>k_1-1}} \bigl ( F(t_2+3,t_2+1) - F(t_2+3,t_2+2) \bigr ) \frac{1}{t_2+2} .
\end{align*}
Thus, taking into account that $F(s,\cdot)$ is a decreasing function (see Lemma  \ref{coro-comparando-div-apendix}), $\Delta \pi_i -\Delta \pi_j\geq 0$, which completes the proof.

\end{proof}

\begin{example}
In transportation networks, the stars, which are known as {\em hub-and-spokes} networks, are specially important. In this framework,  measuring the effect of connecting two different hubs over the centralities of each of the original hubs as well as over the centralities of each of their satellites is a relevant question. Note, that the Position centrality allows us to measure also the importance of the bridge. In this example, we will illustrate the previous result when the functionality of the social network is given by the messages game, which models trips between two nodes of the network.

\begin{table}[h]
    \centering
    \begin{tabular}{c||c|c}
     \hline
     Nodes & $\Delta$ Myerson  &  $\Delta$ Position   
     \\
     \hline
        Hub-$k_1$-star  & $1+\frac{2}{3}(k_1+k_2)+\frac{k_1k_2}{2}$   & $1+k_1+\frac{k_2}{2}+\frac{2k_1k_2}{3}$ \\
        Hub-$k_2$-star & $1+\frac{2}{3}(k_1+k_2)+\frac{k_1k_2}{2}$  &  $1+k_2+\frac{k_1}{2}+\frac{2k_1k_2}{3}$  \\
        Satellites-$k_1$-star &  $\frac{2}{3}+\frac{k_2}{2}$  & $\frac{1}{2} + \frac{k_2}{3}$\\
        Satellites-$k_2$-star &  $\frac{2}{3}+\frac{k_1}{2}$ & $\frac{1}{2} + \frac{k_1}{3}$ \\
       \hline
    \end{tabular}
    \caption{Centralities increments when connecting two hubs.}
    \label{tabla_uniendo_2hubs}
\end{table}

The hub of the larger transportation network gains more Position centrality than the other hub. Moreover, the greater the discrepancy in size between the two networks, the greater the difference in Position centrality gain, which is equal to $\frac{k_2-k_1}{2}$. In terms of satellites, the effect is the opposite. The satellites of the smaller transportation network gain more Position centrality than the other satellites because they are now able to connect to more nodes via the other hub. Again, the greater the discrepancy in size between the two networks, the greater the difference in Position centrality gain, which is $\frac{k_2-k_1}{3}$. Note that Myerson centrality treats both hubs symmetrically, and furthermore they gain less Myerson centrality overall than Position centrality.  The effect on satellite Myerson centrality is similar to position centrality, but in this case satellites gain more Myerson than position centrality.

Additionally, Position centrality allows us to measure the power of connections. The Shapley value of the bridge connecting the two hubs is $2+k_1+k_2+\frac{2k_1k_2}{3}$, which increases with the size of the original hub and spokes transportation networks. As all connections increase their values, it is interesting to compare the value of the bridge in relation to the value of the remaining connections. The ratios of $\phi_b$ over $\phi_{e_1}$ and over $\phi_{e_2}$, where $e_1$ and $e_2$ are spokes of the $k_1$-star and the $k_2$-star, respectively,  are given by:
$$
 \frac{\phi_b}{\phi_{e_1}}=\frac{k_2(1+\frac{2}{3}k_1)+ k_1+2}{\frac{2}{3}k_2+k_1+2}  , \quad \frac{\phi_b}{\phi_{e_2}}=\frac{k_2(1+\frac{2}{3}k_1)+ k_1+2}{k_2+\frac{2}{3}k_1+2} .
$$
Note that even in the limit case with $k_2\to\infty$ the above ratios are greater than one: $\displaystyle\lim_{k_2\to\infty}\frac{\phi_b}{\phi_{e_1}}=\frac{3}{2}+k_1$, and 
 $\displaystyle\lim_{k_2\to\infty}\frac{\phi_b}{\phi_{e_2}}=
1+\frac{2}{3}k_1$, i.e. the value of the bridge remains being strictly greater than the value of any of the original spokes. 

%
%
%
%
%
%
%
%
%
%
%
%

\label{puente-hub-stars-example}
\end{example}

%

\section{The Position attachment centrality}

In this section we analyse the particular case in which the symmetric game is the \emph{attachment game} (Skibski {\it et al.}, 2019) given by $v_a(S)=f_a(s)=2(s-1),\, \forall\, S\subset N$.  Our proposal is to study some specific properties of the \emph{Position attachment centrality} defined, for each graph $\Gamma=(N,E)$ as the position value of the communication situation $(N,v_a,E)$, which we will denote as $PA(\Gamma)$. 

The attachment game was used by Skibski {\it et al.} (2019) to define the graph Attachment centrality, $A(\G)$, as the Myerson value of the attachment game when coalition formation is restricted by a graph $\G$. The Attachment centrality of node $i\in N$ is the expected number of components created by the removal of $i$ (multiplied by $2$ for normalisation purposes) when the nodes are removed from the graph one by one in a random order. The difference between the two centrality measures is due to a difference in the measurement of marginal contributions (see Slikker, 2005). In the case of Attachment centrality, the removal of a node implies the deletion of all its edges simultaneously. On the contrary, when considering the position value, are the edges of the graph which are removed one by one in a random order, and the Position attachment Centrality of $i\in N$ is measured by the sum of the expected number of components created by the removal of each of the edges incident in $i$.

When the graph is a tree, the position value coincides with the Myerson value (see Borm {\it et al.} 1992), and moreover $PA_i(\Gamma)=A_i(\Gamma)=d_i$, where $d_i$ is the degree of node $i$ in $\G$. However, in the general case where the graph contains some cycles, the two centrality measures differ in how end nodes are affected when an edge is removed (or added). According to Attachment centrality, the addition of an edge improves both nodes equally. On the contrary, according to the Position attachment centrality, both nodes are also improved (or at least not worsened), but not necessarily to the same extent. Both attachment centralities also differ on the nodes that are worsened. Formally, let us introduce the concept of {\em intermediary} which is relevant for Position centrality.

Let $\emptyset\neq R\subseteq N$, a subgraph $\Gamma_L=(N[L],L)$ is called a {\em minimal $R$-connecting edge-induced subgraph} (Dietzenbacher {\it et al.}, 2017) if it connects $R$ and any $\Gamma_{L'}$ with $L'\subsetneq L$ does not connect $R$. Let ${\cal M}_\G (R)$ denotes the collection of minimal $R$-connecting edge-induced subgraphs and ${\cal E}_{\Gamma}(R)=\{L_1^R,\dots,L_{q_R}^R\}$ denotes the collection of coalitions of edges which define them. We consider the {\em set of intermediaries of $R$}, which we will denote by $Bet_\G(R)\subseteq N\setminus R$, as the set of nodes in some minimal $R$-connecting edge-induced subgraph. This is:
\begin{equation}
Bet_\G(R):=\{ i\in N\setminus R\tq 
 E_i\cap \Bigl (\cup_{j=1}^{q_R} L^R_j\Bigr ) \neq \emptyset\}.
    \label{interm-R}
\end{equation}

Note that $(N,v_a)\in G^N_0$. Moreover, it is convex. On the contrary, we will show that the corresponding link game $(N,w^a)$ is concave.


\begin{lemma}
   Let $\G=(N,E)$ be a given graph, then the marginal contributions in the link game $(E,w^a)$ defined over the communication situation $(N,v_a,E)$ are given by:
\begin{equation*}
w^a(L\cup \{e\})-w^a(L)=\begin{cases}
0, & \text{if there exists a cycle in $L\cup \{e\}$ containing edge $e$,}
\\
2, &\text{otherwise,}
\end{cases}
\label{CM_at}
\end{equation*}
for every $e\in E$, and every $L\subseteq E\setminus \{e \}$.
 \label{lema_mc_attach}   
\end{lemma}

\begin{proof}
Let $e=\{i,j\}\in E$, and $L\subseteq E\setminus \{e \}$. If there exists a cycle in $L\cup \{e\}$ containing edge $e$ then  $\mathcal{K}(L)=\mathcal{K}(L\cup\{e\})$ and thus $w^a(L\cup \{e\})=w^a(L)$.

Otherwise, there exists $S_i, S_j\in \mathcal{K}(L)$ with $i\in S_i$ and $j\in S_j$ and $S_i\neq S_j$. Thus,
\begin{align*}
    w^a(L) & =  v_a(S_i)+v_a(S_j) +\sum_{\substack{S\in \mathcal{K}(L)\\ S\neq S_i,S_j}} v_a(S) ,
   \\
    w^a(L\cup \{e\})  & =  v_a(S_i\cup S_j) +\sum_{\substack{S\in \mathcal{K}(L)\\ S\neq S_i,S_j}} v_a(S).
\end{align*}
Thus, 
$
w^a(L\cup \{e\})-w^a(L)=v_a(S_i\cup S_j)-v_a(S_i)-v_a(S_j)=2.
$
\end{proof}

Note that the marginal contributions in the link game are non-increasing, and thus $(E,w^a)$ is concave for every given graph $\Gamma=(N,E)$.

\begin{proposition}
Let $\G=(N,E)$ be a given graph. If an edge $e_0=(i_0,j_0)$, between two nodes $i_0,j_0\in N$ is added. Then, it holds that:
\begin{enumerate}[$(i)$]
\item $PA_i(N,E\cup\{ e_0\})\leq PA_{i}(N,E)$, for every intermediary node $i\in Bet(\{i_0,j_0\})$.
\item $PA_i(N,E\cup\{ e_0\})= PA_{i}(N,E)$, for every  node $i\notin Bet(\{i_0,j_0\})\cup \{i_0,j_0\}$
\item $PA_{i_0}(N,E\cup\{ e_0\})\geq PA_{i_0}(N,E)$ and $PA_{j_0}(N,E\cup\{ e_0\})\geq PA_{j_0}(N,E)$
\end{enumerate}
\end{proposition} 

\begin{proof}
Taking into account Lemma \ref{lema_mc_attach},   for every edge $e\in E$, and every $L\subseteq E\setminus \{e\}$, it is verified:
\begin{equation}
w^a(L\cup \{e,e_0\})-w^a(L\cup\{ e_0\})  \leq w^a(L\cup \{e\})-w^a(L).
\label{MCe-des}
\end{equation}
Then, for every edge $e \in E$ it holds: 
\begin{multline*}
\phi_{e} (E\cup \{e_0\},w^a)=\sum_{L\subseteq E\setminus \{ e\}} \frac{l ! (m-l)!}{(m+1)!}(w^a(L\cup \{e\})-w^a(L)) + 
\\
 + \frac{(l+1) ! (m-l-1)!}{(m+1)!}(w^a(L\cup \{e,e_0\})-w^a(L\cup\{ e_0\}) ) \leq 
 \\
 \leq \sum_{L\subseteq E\setminus \{ e\}} \frac{l ! (m-l-1)!}{m!}(w^a(L\cup \{e\})-w^a(L))=
 \phi_{e} (E,w^a) ,
\end{multline*}
where $m=\vert E\vert$. Therefore, $PA_i(N,E\cup\{ e_0\})\leq PA_{i}(N,E)$, for every node $i\in N\setminus\{i_0,j_0\}$, since for the end nodes $i_0,j_0$, $\frac{1}{2}\phi_{e_0}(E\cup \{e_0\},w^a)$ must be added.

 Now, in order to prove statement $(ii)$, let $e \notin \cup_{j=1}^{q_0} L_j^0$, being ${\cal E}_\G (\{i_0,j_0\})=\{L_1^0,\dots,L_{q_0}^0\}$. Then, there exists a cycle containing edge $e$ in $L\cup \{e,e_0\}$ if, and only if, there exists a cycle containing edge $e$ in $L\cup\{e\}$. Thus, inequality \eqref{MCe-des} is an equality for every subset $L\subseteq E\setminus \{ e\}$, and therefore  $\phi_{e} (E,w^a) = \phi_{e} (E\cup\{e_0\},w^a)$. If $i\notin Bet(\{i_0,j_0\})\cup \{i_0,j_0\}$, then $E_i(N,E\cup\{ e_0\})=E_i(N,E)\subseteq E\setminus \cup_{j=1}^{q_0} L_j^0$ and therefore $PA_i(N,E\cup\{ e_0\})= PA_{i}(N,E)$.

We will restrict the proof of $(iii)$ to the case of connected graphs. Note that for not connected graphs, two cases are possible:
\begin{enumerate}
    \item $i_0$ and $j_0$ belong to different connected components. Then, since the attachment game is convex, by Proposition \ref{adding-bridge-prop}, $(iii)$ holds.
    \item $i_0$ and $j_0$ belong to the same connected component, then taking into account that Position centrality verifies Locality (see Proposition \ref{locality}), we can restrict the proof to the connected subgraph induced by their connected component.
    \end{enumerate}

The proof relies on the alternative characterisation of the Shapley given by Weber (1988) in terms of all possible orders of arrival of the players to a meeting point. We will prove the inequality for $i_0$, the same reasoning applies to $j_0$.

Let $\theta\in \Theta^m$ be a given order of the original edges in $E$.  Then, it holds:
\begin{equation}
 PA_{i_0} (N,E) =\frac{1}{2\cdot m!} \sum_{\theta \in \Theta^m}
  \sum_{e \in E_{i_0}} \Bigl ( w^a (Pred^{e} (\theta) \cup \{e\})-w^a (Pred^{e} (\theta)) \Bigr ),
  \label{order-original}
  \end{equation}
  where $Pred^{e} (\theta)\subseteq E\setminus \{ e\}$ denotes the set of edges that precede edge $e$ in the order $\theta$. 

  Now, let $\theta \in \Theta^m$ be any given order of the original edges,  we will prove that:
  \begin{eqnarray}
     \sum_{e \in E_{i_0}} \Bigl ( w^a (Pred^{e} (\theta) \cup \{e\})-w^a (Pred^{e} (\theta)) \Bigr ) \leq \hspace*{6cm}
 \label{sumando-theta}
   \\
   \sum_{e \in E_{i_0}} \bigl ( w^a (Pred^{e} (\theta^+_k) \cup \{e\})-w^a (Pred^{e} (\theta^+_k) \bigr ) 
   + \bigl ( w^a (Pred^{e_0} (\theta^+_k) \cup \{e_0\})-w^a (Pred^{e_0} (\theta^+_k) \bigr )  , 
   \label{sumando-theta-k}
  \end{eqnarray}
  for each of the $m+1$ orders of the edges in $E\cup \{e_0\}$, $\theta^{+}_k\in \Theta^{m+1}$, defined by means of inserting edge $e_0$ in every possible position $k$, $1\leq k\leq m+1$, of the given order $\theta$.
  
  %
%
Making use of Lemma \ref{lema_mc_attach}, we only need to prove that for every given order $\theta \in \Theta^m$, the number of edges with a nonzero contribution in \eqref{sumando-theta} is less or equal than the number of edges with a nonzero contribution in \eqref{sumando-theta-k}, for every $1\leq k\leq m+1$. For doing it, we will define a partition of the collection of paths, ${\cal M}_\G (\{i_0,j_0\})$, by means of their starting edge from $i_0$. Formally, let $E_{i_0} (j_0)\subseteq E_{i_0}$ be the collection of edges incident in $i_0$ that form part of some of these  paths in ${\cal M}_\G (\{i_0,j_0\})$. Then, we define the partition ${\cal P} (e):=\{ P\in {\cal M}_\G (\{i_0,j_0\})\, / \,  e \in P\}$, for each $e \in E_{i_0} (j_0)$. 
 

Let $\theta \in \Theta^m$ be a given order of the original edges, and let $e\in E_{i_0}$ with $w^a(Pred^{e} (\theta) \cup \{e\})-w^a (Pred^{e} (\theta)) =2$. Then:
\begin{enumerate}
    \item For every position $k>\theta(e)$ it holds:
    $$
    w^a (Pred^{e} (\theta) \cup \{e\})-w^a (Pred^{e} (\theta)) =
    w^a (Pred^{e} (\theta^+_k) \cup \{e\})-w^a (Pred^{e} (\theta^+_k)) .
    $$
    \item Analogously, if $k\leq \theta(e)$, but $e\notin E_{i_0}(j_0)$, i.e. $e\notin \cup_{j=1}^{q_0} L_j^0$, then:
    $$
    w^a (Pred^{e} (\theta) \cup \{e\})-w^a (Pred^{e} (\theta)) =
    w^a (Pred^{e} (\theta^+_k) \cup \{e\})-w^a (Pred^{e} (\theta^+_k)) .
    $$
    \item If $k\leq \theta(e)$ and $e\in E_{i_0}(j_0)$. Then, one of the following three conditions -which are mutually exclusive- is verified:
  \begin{itemize}
      \item[(c1)] There exists a path in $P\in {\cal P} (e)$ such that $P\setminus \{ e\} \subseteq Pred^{e} (\theta)$. 
      \item[(c2)] There exists $P\in {\cal P} (e')$, $e'\in E_{i_0} (j_0)\setminus \{e\}$, with $P\subseteq Pred^{e} (\theta)$.
      \item[(c3)] None of the two previous conditions hold.
  \end{itemize}

If (c1) holds, then $w^a (Pred^{e} (\theta^+_k) \cup \{e\})-w^a(Pred^{e} (\theta^+_k)) =0$.   However, since condition (c2) is not true, then it was also not true when edge $e_0$ arrived, and therefore $w^a (Pred^{e_0} (\theta^+_k) \cup \{e_0\})-w^a (Pred^{e_0} (\theta^+_k) )=2$. Thus, $e_0$ takes the place of $e$ in \eqref{sumando-theta-k}. Note that there is a unique edge $e$ in this situation per each order $\theta\in \Theta^{m}$.     

On the other hand, if (c2) holds there exists a path from $i_0$ to $j_0$ in $Pred^{e} (\theta)$, and therefore the arrival of edge $e_0$ does not affect the marginal contribution of $e$. Trivially, if (c3) holds, again the arrival of edge $e_0$ does not affect the marginal contribution of $e$. Thus, in both cases, its is verified:
$$
    w^a (Pred^{e} (\theta) \cup \{e\})-w^a (Pred^{e} (\theta)) =
    w^a (Pred^{e} (\theta^+_k) \cup \{e\})-w^a (Pred^{e} (\theta^+_k)) .
    $$
\end{enumerate}
To sum up, it is satisfied:
\begin{multline*}
  PA_{i_0} (N,E\cup\{e_0\}) = 
  \frac{1}{2\cdot(m+1)!} \sum_{\theta \in \Theta^{m+1}}
\Biggl [ \bigl ( w^a (Pred^{e_0} (\theta) \cup \{e_0\})-w^a(Pred^{e_0} (\theta) \bigr ) +
  \\
  + \sum_{e \in E_{i_0}} \bigl ( w^a (Pred^{e} (\theta) \cup \{e\})-w^a (Pred^{e} (\theta) \bigr ) \Biggr ] =
   \\ 
  \frac{1}{2\cdot (m+1)!} \sum_{\theta \in \Theta^{m}} \sum_{k=1}^{m+1} 
   \Biggl [ \bigl ( w^a (Pred^{e_0} (\theta^+_k) \cup \{e_0\})-w^a (Pred^{e_0} (\theta^+_k) \bigr )  
   \\
   + \sum_{e \in E_{i_0}} \bigl ( w^a (Pred^{e} (\theta^+_k) \cup \{e\})-w^a (Pred^{e} (\theta^+_k) \bigr )\Biggr ] \geq 
   \\
   \geq 
   \frac{1}{2\cdot (m+1)!} \sum_{\theta \in \Theta^m}
  (m+1)\sum_{e \in E_{i_0}} \Bigl ( w^a (Pred^{e} (\theta) \cup \{e\})-w^a (Pred^{e} (\theta)) \Bigr )=PA_{i_0}(N,E),
\end{multline*}
and statement $(iii)$ holds.
\end{proof}

\begin{example}
To analyze the difference between both centralities   measures based on attachment we consider the addition of edge $\{2,15\}$ in the graph of Example \ref{example_functionality_M}:

\vspace*{2mm}
\begin{center}

\begin{tikzpicture}[node_style/.style={draw,circle,minimum size=0.5cm,inner sep=1}]

\node[node_style] (siete) at (0,1.2) {7} ;
\node[node_style] (seis) at (0.75,2.4) {6} ;
\node[node_style] (ocho) at (0.75,0) {8} ;
\node[node_style] (cuatro) at (1.5,1.2) {4} ;
\node[node_style] (cinco) at (2.25,2.4) {5} ;
\node[node_style] (nueve) at (2.25,0) {9} ;
\node[node_style] (uno) at (3,1.2) {1} ;
\node[node_style] (dos) at (4.5,1.2) {2} ;
\node[node_style] (tres) at (6,1.2) {3} ;
\node[node_style] (doce) at (9,1.2) {12} ;
\node[node_style] (once) at (8.25,2.4) {11} ;
\node[node_style] (trece) at (8.25,0) {13} ;
\node[node_style] (quince) at (7.5,1.2) {15} ;
\node[node_style] (diez) at (6.75,2.4) {10} ;
\node[node_style] (catorce) at (6.75,0) {14} ;

\draw (uno) -- (cinco);
\draw (uno) -- (nueve);
\draw (uno) -- (dos);
\draw (dos) -- (tres);
\draw (cuatro) -- (cinco);
\draw (cuatro) -- (seis);
\draw (cuatro) -- (siete);
\draw (cuatro) -- (ocho);
\draw (cuatro) -- (nueve);
\draw (cinco) -- (seis);
\draw (seis) -- (siete);
\draw (siete) -- (ocho);
\draw (ocho) -- (nueve);

\draw (quince) -- (diez);
\draw (quince) -- (once);
\draw (quince) -- (doce);
\draw (quince) -- (trece);
\draw (quince) -- (catorce);
\draw (diez) -- (once);
\draw (once) -- (doce);
\draw (doce) -- (trece);
\draw (trece) -- (catorce);
\draw (tres) -- (diez);
\draw (tres) -- (catorce);

\draw [dashed,line width=1.0pt] (dos) edge[-,bend right=30]  (quince);
\end{tikzpicture}
    
\end{center}

In Table \ref{example_comparando_MAyPA} are depicted the centrality variation of each node. Note that node 15 gains more Position attachment centrality than the other end node 2 ($0.56$ versus $0.26$), whereas both gain the same attachment centrality. With respect to the remaining nodes, only the attachment centrality of the bypassed node 3, and that of the intermediary adjacent nodes  10 and 14, which were in the shortest paths between 2 and 15, is reduced. However, according to Position attachment centrality,  also nodes 11, 12 and 13, which were also potential intermediaries between 2 and 15 although but they did not appear in any shortest path, reduce their centrality. Myerson centrality only affects to nodes in minimal paths with respect to nodes, whereas Position centrality affects to nodes in minimal paths with respect to edges.

If we consider a local centrality measure such as degree, the addition of the edge $\{2,15\}$ affects both end nodes equally, but if we consider a more global measure such as closeness, node 15 also gains more closeness centrality than node 2. 

\begin{table}[h]
    \centering
    \begin{tabular}{c|c|c}
     \hline
     Nodes & $\Delta$ Attachment &  $\Delta$  Position Attachment 
     \\
     \hline
        2 & 0.4   & 0.26 
        \\
        15 & 0.4  & 0.56 
        \\
        3 &  -0.6  & -0.52 
        \\
        10,14 & -0.1  & -0.12 
        \\
        11,13 & 0 & -0.03  
        \\
         12 &   0 & -0.01 
         \\
         1,4,\dots,9 &  0 & 0 
         \\
        \hline
    \end{tabular}
    \caption{Attachment centralities increments}
    \label{example_comparando_MAyPA}
\end{table}

\end{example}

\subsection{A characterisation of the Position attachment centrality}

The characterisation of the Position attachment centrality is based on Position value (Slikker, 2005) and attachment centrality (Skibski {\it et al.}, 2019) characterisations. For this purpose, the following axioms for a centrality measure $\sigma$ are considered.



\vspace{0.2cm}
\noindent {\bf Locality Axiom} 

For every graph $\Gamma = (N, E)$ and every node $i \in N$ , the
centrality of $i$ depends solely on the component to which $i$ belongs, i.e. $\sigma_i(\Gamma)=\sigma_i((\K_i(\Gamma), E[\K_i(\Gamma)]))$.

\vspace{0.2cm}
\noindent {\bf Normalisation Axiom} 

\begin{itemize}
    \item $\sigma_i(\Gamma) \in [0, n-1]$;
    \item $\sigma_i(\Gamma)=0$ when $i$ is isolated in $\Gamma$;
    \item $\sigma_1(\Gamma^*)= n-1$ when $\G^*$ is a star with $n$ nodes and being $1$ its hub.
\end{itemize}

\noindent \vspace{0.2cm}
{\bf Gain-Loss Axiom} 

For every connected graph, $\G = (N,E)$, and every
pair of nodes, $i,j \in N$ , adding the edge $\{i,j\}\notin E$ to $E$ does not affect the sum of centralities.

\vspace{0.2cm}
\noindent  {\bf Balanced link contributions Axiom}

For every $\G=(N, E)$ and all $i,j\in N$
\begin{equation*}
\sum_{l\in E_{j}}\left( \sigma_i(\Gamma) -\sigma_i(\Gamma- l)\right) =\sum_{l\in E_{i}}\left( \sigma_j(\Gamma) -\sigma_j(\Gamma- l)\right) \text{,}
\end{equation*}
where $\G-l:=(N,E\setminus\{ l\})$.

\begin{theorem}
The Position attachment centrality is the unique centrality measure that satisfies Locality, Normalisation, Gain-Loss and Balanced link contributions.
\end{theorem}

\begin{proof}
As a position value $PA(\G)$ satisfies Locality (proposition \ref{locality}) and Balanced link contributions (Slikker, 2005). Moreover, by proposition \ref{ComponentEfficiency}, for all connected component $C\in K(\G)$
\begin{equation*}
    \sum_{i\in C} PA_i(\G)=2(|C|-1). 
\end{equation*}

Then, adding a new edge to a connected component does not affect the sum of centralities of its nodes, thus verifying Gain-Loss. 

With respect to Normalisation, by proposition \ref{starmax}, the node hub $1$ of a star of $n$ nodes verifies that $\pi_1(N,v_{a},E^*)=f_{a}(n)/2$. Then, $PA_1=n-1$. By proposition \ref{isolatedmin}, for an isolated node $k$, $PA_k=0$. Finally, by propositions \ref{positivity} and \ref{starmax}, $0\leq PA_i(\G)\leq n-1, \forall i\in N$ holds. 

For proving uniqueness, assume that $\sigma(\G)$ is a centrality measure that satisfies the four axioms above. First we show that for all $C\in K(\G)$, 
$$
\sum_{i\in C} \sigma_i(\G)= 2(|C|-1).
$$

Consider a $n$ nodes star $\G^*$, being node $1$ its hub. By Normalisation  $\sigma_1(\G^*)= n-1$ and $\sigma_i(\G^* - \{1,i\})=0$ for all $i\neq 1$, and from Locality and Normalisation $\sigma_1(\G^* - \{1,i\})=n-2$ for all $i\in N\setminus\{1\}$.

Using Balanced link contributions we will show by induction on $n$ that $\sigma_i(\G^*)=1$ for all $i\in N\setminus\{1\}$. For $n=3$:
$$
2\sigma_i(\G^*)-\sigma_i(\G^* -l_i)-\sigma_i(\G^* - l_j)=\sigma_1(\G^*)-\sigma_1(\G^* - l_i)=1
$$
and as $\sigma_i(\G^*- l_i)=0$ and $\sigma_i(\G^* -l_j)=1$, $\sigma_i(\G^*)=1$.

Assume that, $\sigma_i(\G^*)=1$ for all node $i\neq 1$ of a star with $n-1$ nodes. Then, by Balanced link contributions:
$$
\sum_{l_j\in L_1} [\sigma_i(\G^*)-\sigma_i(\G^*- l_j)]=\sigma_1(\G^*)-\sigma_1(\G^*- l_i )=1
$$

But, by induction hypothesis, Locality and Normalisation $\sigma_i(\G^* - l_i)=0$ and $\sigma_i(\G^*- l_j)=1$ for all $j\neq i$, then $\sigma_i(\G^*)=1$.

Then $\sum_{i\in N} \sigma_i(\G^*)= 2(n-1)$. Taking into account that any connected graph $\G$ with $n$ nodes can be obtained adding or removing edges from a star and Gain-Loss, we obtain that  $\sum_{i\in N} \sigma_i(\G)=2(n-1)$. Finally, by Locality, $\sum_{i\in S} \sigma_i(\G)= 2(|S|-1)$ for all connected component $S\subset N$ of any graph $\G$.

Therefore, $\sigma(\G)$ verifies Component efficiency for the game $f(S)=2(|S|-1)$ and Balanced link contributions. By Slikker (2005) characterisation, $\sigma(\G)=PA(\G)$ for all $\G$.

\end{proof}


\section{Some conclusions}
\label{Conclusions}


In this work, we have proposed a family of node centrality measures based on the evaluation of its links using the position value. We have shown that this family, under certain conditions, this family satisfies  typical properties of a centrality measure. In particular, it verifies 
component efficiency and it assigns a minimum value to the isolated nodes. It also ensures that, among all graphs with $n$ nodes, maximum centrality is achieved by the hub of a star and minimum centrality is achieved  by the end nodes of a chain. In the case of chain, centrality  increases progressively from the end nodes towards the median nodes.

Compared to centrality measures based on the Myerson value, our approach relaxes the fairness condition, which requires that the removal/addition of a link affects equally its incident nodes. Our proposal is thus more realistic, as proved in the case of two
hub-and-spokes networks connected by their hubs. In this particular example, we have shown that the hub of the larger transportation network gains more position centrality than the other hub. However, we note  the limitation of our approach to determine the position centrality variation of two nodes once a bridge is  established between them in arbitrary communication situations. Concretely, it could be interesting to describe and explain situations where one of the bridge nodes loses centrality. 
In the particular case of communication situations where the game is the attachment game, it is shown that the two incident nodes improve their centrality.

Finally, based on the work of Skibski {\it et al.} (2019) on attachment centrality, we have characterised the Position attachment centrality according to four axioms:  locality, normalisation, gain-loss and balanced link contributions. It would be interesting to extend this specific analysis to other prominent members of the family.

In order to demonstrate some of the above results, compelling properties of the Harsanyi dividends for the link game, which are interesting in themselves, have been shown.

\section*{Appendix}

\setcounter{lemma}{1}
\addtocounter{lemma}{-1}
\begin{lemma}
Let $\left( N,v\right) $ be a $TU$ game and $\Gamma =\left( N,E\right) \in 
\mathcal{G}^{N}$. For every subset $L\subseteq E$ such that $\Gamma_L$ is not connected it holds $\lambda_{L}\left(w^{v}\right) = 0$.
\end{lemma}

\begin{proof}
If $\Gamma _{L}$ is not connected, $\mathcal{K(}L)$ has at least two
elements. W.l.o.g. suppose that $\K(L)=\{S_1, S_2\}$. Let $L_{r}=E[S_r]$, for $r=1,2$. Then, $L_{1}\cap L_{2}=\varnothing$, and 
\begin{equation*}
w^{v}\left( L\right) =w^{v}\left( L_{1}\right) +w^{v}\left( L_{2}\right) .
\end{equation*}%
Note that each subset $T\subseteq L$ can be decomposed as $T_{1}\cup T_{2}$
where $T_{r}\subseteq L_{r},r=1,2$. Let us calculate the Harsanyi dividend
of $L$:
\begin{align*}
\lambda_{L}\left( w^{v}\right)  &=\sum_{T\subseteq L}\left(
-1\right) ^{l-t}w^{v}\left( T\right) =\sum_{\substack{ T_{1}\subseteq L_{1}
\\ T_{2}\subseteq L_{2}}}\left( -1\right) ^{l-t_{1}-t_{2}}w^{v}\left(
T_{1}\cup T_{2}\right)  \\
&=\sum_{\substack{ T_{1}\subseteq L_{1} \\ T_{2}\subseteq L_{2}}}\left(
-1\right) ^{l-t_{1}-t_{2}}\left( w^{v}\left( T_{1}\right) +w^{v}\left(
T_{2}\right) \right)  \\
&=\sum_{_{\substack{ T_{1}\subseteq L_{1} \\ T_{2}\subseteq L_{2}}}}\left(
-1\right) ^{l-t_{1}-t_{2}}w^{v}\left( T_{1}\right) +\sum_{_{\substack{ %
T_{1}\subseteq L_{1} \\ T_{2}\subseteq L_{2}}}}\left( -1\right)
^{l-t_{1}-t_{2}}w^{v}\left( T_{2}\right) \text{.}
\end{align*}%
Note that the factor $w^{v}\left( T_{1}\right) $ appears in the summation as
many times as subsets of edges of $L_{2} $, and analogously occurs with $
w^{v}\left( T_{2}\right) $. Then%
\begin{align*}
\lambda_{L}\left( w^{v}\right)  &=\sum_{T_{1}\subseteq
L_{1}}\left( -1\right) ^{l-t_{1}}w^{v}\left( T_{1}\right) \left(
\sum_{T_{2}\subseteq L_{2}}\left( -1\right) ^{-t_{2}}\right)
+\sum_{T_{2}\subseteq L_{2}}\left( -1\right) ^{l-t_{2}}w^{v}\left(
T_{2}\right) \left( \sum_{T_{1}\subseteq L_{1}}\left( -1\right)
^{-t_{1}}\right) 
\\
&=\sum_{T_{1}\subseteq L_{1}}\left( -1\right) ^{l-t_{1}}w^{v}\left(
T_{1}\right) \left( \sum_{k=0}^{l_{2}}\binom{l_{2}}{k}\left( -1\right)
^{-k}\right) +\sum_{T_{2}\subseteq L_{2}}\left( -1\right)
^{l-t_{2}}w_{E}^{v}\left( T_{2}\right) \left( \sum_{k=0}^{l_{1}}\binom{l_{1}%
}{k}\left( -1\right) ^{-k}\right) \\
&=\sum_{T_{1}\subseteq L_{1}}\left( -1\right) ^{l-t_{1}}w^{v}\left(
T_{1}\right) \left( 1-1\right) ^{l_{1}}+\sum_{T_{2}\subseteq L_{2}}\left(
-1\right) ^{l-t_{2}}w^{v}\left( T_{2}\right) \left( 1-1\right) ^{l_{2}}=0%
\text{.}
\end{align*}
\end{proof}

\setcounter{lemma}{4}
\begin{lemma}
Let $(N,v)\in G^N_0$ and $\Gamma =\left( N,E\right) \in \mathcal{G}^{N}$. If $\Gamma$  is cycle-free and $L\subseteq E$,\thinspace $L\neq \varnothing $, is connected, then the Harsanyi dividend of $L$ in
the link game $w^{v}$ can be computed as follows:%
\begin{equation}
\lambda _{L}\left( w^{v}\right) =\sum_{\mathcal{D}\left( L\right) \subseteq
T\subseteq L}\left( -1\right) ^{l-t}w^{v}\left( T\right) 
\label{HarsayiD(L)}
\end{equation}
where $\mathcal{D}\left( L\right) $ is the cut edges set of $L$.
\end{lemma}

\begin{proof}
First, we decomposed the sum in \eqref{harsanyi-div-def} as follows:%
\begin{equation*}
\lambda _{L}\left( w^{v}\right) =\sum_{T\subseteq L}\left( -1\right)
^{l-t}w^{v}\left( T\right) =\sum_{\mathcal{D}\left( L\right) \subseteq
T\subseteq L}\left( -1\right) ^{l-t}w^{v}\left( T\right) +\sum_{T\in \T (L)}\left( L\right) \left( -1\right) ^{l-t}w^{v}\left( T\right) ,
\end{equation*}%
where $\mathcal{T}\left( L\right) =\left\{ T\subseteq L\mid \mathcal{D}%
\left( L\right) \nsubseteq T\right\} $. Then, we prove that the second summand  is zero. 

If $\D(L)=\varnothing$ or $\T(L)=\varnothing$, then trivially \eqref{HarsayiD(L)} holds. Otherwise, suppose  $\mathcal{D}\left( L\right) \neq \varnothing  $ and $\mathcal{T}\left( L\right) \neq  \varnothing  
$. Due to the definition of the link game we only have to consider connected subsets $T\subseteq \T(L)$, since for a non-connected $T$ its link game value would be the sum of the link game values corresponding to its components.  


$w^v(T)$ will appear in the summands corresponding to all $T'$ such that $T\subseteq T'\subseteq \T(L)$ and being $T$ one of its connected components. The number of subsets $T'$ with this property of size $t+m$ is 
$$\binom{l-d\left(
T\right) -t}{m},
$$
where $d(T)$ is the number of adjacent links to nodes in $N[T]$ that are not in $T$. Thus, the coefficient of $w^v(T)$ is giving by:%
\begin{equation*}
\sum_{m=0}^{l-d\left( T\right) -t}\left( -1\right) ^{l-t-m}\binom{l-d\left(
T\right) -t}{m} =\left( -1\right) ^{l-t}\left( 1-1\right) ^{l-d\left(
T\right) -t},
\end{equation*}%
which equals 0, since there is at least one subset $T^{\prime }\in \mathcal{T}\left( L\right)$,
such that $T^{\prime }$ is not connected and $T\subset T^{\prime }$ is one of its connected components.

%
%
\end{proof}

\setcounter{lemma}{1}
\begin{lemma}
Let $(N,v)\in G^N_0$ and $\Gamma =\left( N,E\right) \in \mathcal{G}^{N}$. If $\Gamma$  is cycle-free, then for all $\emptyset \neq L\subseteq E$ with $\Gamma_L$ connected, the Harsanyi dividends of the link game can
be calculated in the following way:%
\setcounter{equation}{8}
\begin{equation}
\lambda _{L}\left( w^{v}\right) =F\left( l+1,l-d_{L}\right),
\end{equation}%
where $d_{L}=\left\vert \mathcal{D}\left( L\right)\right\vert $, being $F(s,r)= \sum_{k=0}^{r}\left( -1\right) ^{k}\binom{r}{k}f\left(s-k\right)$, for $s, r\in \mathbb{N}$ and $s\geq r$.
\end{lemma}

\begin{proof}
Let $\emptyset \neq L\subseteq E$ with $\Gamma_L$ connected, then by \eqref{HarsayiD(L)}, and taking into account the symmetry of $v$, it holds:
\setcounter{equation}{23}
\begin{align}
\lambda _{L}\left( w^{v}\right) & =\sum_{\mathcal{D}\left( L\right) \subseteq
T\subseteq L}\left( -1\right) ^{l-t}w^{v}\left( T\right) =\sum_{\mathcal{D} \left( L\right) \subseteq T\subseteq L}\left( -1\right) ^{l-t}f\left(
t+1\right)
\label{divF1}
\\
&  =\sum_{m=0}^{l-d_{L}}\left( -1\right) ^{l-d_{L}-m}%
\binom{l-d_{L}}{m}f\left( d_{L}+m+1\right)=F\left( l+1,l-d_{L}\right).
\label{divF2}
\end{align}%

\end{proof}

\setcounter{corollary}{1}
\begin{corollary}
Let $(N,v)\in G^N_0$ and $\Gamma =\left( N,E\right) \in \mathcal{G}^{N}$. If $\Gamma$  is cycle-free, $f\in \mathcal{C}^{n}\left( \mathbb{R}\right) $ and $f^{\left(
k\right) }\left( x\right) \geq 0$ in $[1,+\infty)$, for  $k=0,1,...,n$, then $\lambda _{L}\left( w^{v}\right) \geq 0$, for all $\emptyset \neq L\subseteq E$.
\end{corollary}

\begin{proof}
The expression of the dividends given by \eqref{divF2} is a numerical 
approximation of the derivative of $f$ of order $d_{L}+1\leq n,$ and we know that $%
f^{\left( d_{L}+1\right) }\left( x\right) \geq 0$ in a intermediate point $%
x\in \lbrack d_{L}+1,l+1]$, then $\lambda _{L}\left( w^{v}\right) \geq 0$.
\end{proof}

\begin{lemma}
Let $f$ be a real function such that $f^{\left(
k\right) }\left( x\right) \geq 0$ in $[1,+\infty)$, for  $k=0,1,...,n$, then  the function $F(s,r)$ is decreasing in the second argument for every fixed $s\in \mathbb{N}$, with $r\in \mathbb{N}, s\geq r$.
\label{coro-comparando-div-apendix}
\end{lemma}

\begin{proof}
We first prove that:%
\begin{equation*}
F\left( s,r\right) +F\left( s-1,r-1\right) =F\left( s,r-1\right) \text{%
,}
\end{equation*}%
where $1<r<s.$ From definition%
\begin{align*}
F\left( s,r-1\right)  &=\sum_{k=0}^{r-1}\left( -1\right) ^{k}\binom{r-1}{k%
}f\left( s-k\right)  
\\
&=F\left( s,r\right) -\left( -1\right)^{r}f\left( s-r\right)
-\sum_{k=0}^{r-1}\left( -1\right)^{k}\left[ \binom{r}{k}-\binom{r-1}{k}%
\right] f\left( s-k\right)  
\\
&= F\left( s,r\right) +\left( -1\right)
^{r-1}f\left( s-r\right) +\sum_{k=1}^{r-1}\left( -1\right)^{k-1}\binom{r-1}{%
k-1}f\left( s-k\right) 
\\
&=F\left( s,r\right) +\sum_{m=0}^{r-1}\left( -1\right)^{m-1}\binom{r-1}{m}f\left( s-1-m\right) =F\left( s,r\right) +F\left( s-1,r-1\right) .
\end{align*}

%
%
%
Then, since $F\left( s-1,r-1\right) \geq 0$ by Corollary \ref{dividendo-positivo-fcond}, we have $F\left( s,r\right) \leq F\left( s,r-1\right)$. Thus recursively, we obtain 
$
F\left( s,r\right) \leq F\left( s,r^{\prime }\right), $ 
whenever $r^{\prime }\leq r$, and the results holds. 
\end{proof}


Next corollary follows straightforwardly from Lemmas \ref{dividendos-exp-F} and \ref{coro-comparando-div-apendix}. 

\setcounter{corollary}{1}
\addtocounter{corollary}{-1}
\begin{corollary}
Let $(N,v)\in G^N_0$ and $\Gamma =\left( N,E\right) \in \mathcal{G}^{N}$. If $\Gamma$  is cycle-free, $f\in \mathcal{C}^{n}\left( \mathbb{R}\right) $ and $f^{\left(
k\right) }\left( x\right) \geq 0$ in $[1,+\infty)$, for  $k=0,1,...,n$, then, for every pair of subsets $L,L^{\prime }\subseteq E$, such that $l=l^{\prime }$ and $d_{L}\leq d_{L^{\prime }}$ it holds:
$$
 \lambda_{L}(w^{v})\leq \lambda_{L'}(w^{v})
 $$
\end{corollary}


\begin{thebibliography}{}

\bibitem{BaMuNa17} Bandyopadhyay, S., Murty, M. N., \& Narayanam, R. (2017). A generic axiomatic characterization of centrality measures in social network. arXiv preprint arXiv:1703.07580.

\bibitem{BlJaTe16} Bloch, F., Jackson, M. O. and Tebaldi, P. (2021). Centrality measures in networks. {\it arXiv preprint}, arXiv:1608.05845.

\bibitem{BlJaTe21} Bloch, F., Jackson, M. O., \& Tebaldi, P. (2016). Centrality measures in networks. arXiv preprint arXiv:1608.05845.

\bibitem{BoVi14} Boldi, P., \& Vigna, S. (2014). Axioms for centrality. Internet Mathematics, 10(3-4), 222-262.

\bibitem{Bo87} Bonacich, P. (1987). Power and centrality: A family of measures. {\it American journal of sociology} 92(5), 1170-1182.

\bibitem{Bo03} Borgatti, S. P. (2003). {\it Identifying sets of key players in a network}. In: IEMC'03 Proceedings. Managing Technologically Driven Organizations: The Human Side of Innovation and Change (IEEE Cat. No. 03CH37502), 127-131, IEEE.

\bibitem{BoEv06} Borgatti, S. P. and Everett, M. G. (2006). A graph-theoretic perspective on centrality. {\it Social Networks} 28(4), 466-484.

\bibitem{BoOwTi92} Borm, P., Owen, G., Tijs, S.H. (1992). On the position value for
communication situations. {\it SIAM J. Discrete Math.} 5, 305-320.



\bibitem{DiBoHe17} Dietzenbacher, B., Borm, P., Hendrickx R. (2017).  Decomposition of network communication games. {\it Mathematical Methods of Operations Research} 85: 407-423.

\bibitem{Est06} Estrada, E. (2006). Virtual identification of essential proteins within the protein interaction network of yeast. {\it Proteomics}, 6(1), 35-40.

\bibitem{Fr79} Freeman, L.C. (1979). {Centrality in social networks: conceptual clarification}. {\it Social Networks} 1, 215-239.


\bibitem{Fr91} Friedkin, N. E. (1991). Theoretical foundations for centrality measures. {\it American Journal of Sociology} 96(6), 1478-1504.


\bibitem{Ga09} Garg, M. (2009). Axiomatic foundations of centrality in networks. Available at SSRN 1372441.

\bibitem{GiNe02} Girvan, M. and Newman, M. E. J. (2002). Community structure in social and biological networks. {\it Proc. Natl Acad. Sci. USA 99}, 7821-7826.

\bibitem{GoGoMaPoTe03} G\'{o}mez, D., Gonz\'{a}lez-Arang\"{u}ena, E., Manuel, C., Owen, G., del Pozo, M. and Tejada, J. (2003). Centrality and power in social networks: a game theoretic approach. {\it Mathematical Social Sciences} 46, 27-54.

\bibitem{GoGoMaOwPo04} G\'{o}mez, D., Gonz\'{a}lez-Arang\"{u}ena, E., Manuel, C., Owen, G., \& del Pozo, M. (2004). A unified approach to the Myerson value and the position value. Essays in Cooperative Games: In Honor of Guillermo Owen, 63-76.

\bibitem{Go17} Gofman, M. (2017). Efficiency and stability of a financial architecture with too-interconnected-to-fail institutions. {\it Journal of Financial Economics} 124(1), 113-146.

\bibitem{GrOw82} Grofman, B., Owen, G. (1982). A game-theoretic approach to measuring centrality in social networks. {\it Social Networks} 4, 213-224.

\bibitem{HaMeKu16} Hajian, M., Melachrinoudis, E., \& Kubat, P. (2016). Modeling wildfire propagation with the stochastic shortest path: A fast simulation approach. {\it Environmental modelling \& software}, 82, 73-88.

\bibitem{Ha59} Harsanyi, J.C. A bargaining model for cooperative n-person games. In {\it Contributions to the Theory of Games IV}; Tucker, A.W., Luce,
R.D., Eds.; Priceton University Press: Princeton, NJ, USA, 1959; pp. 325-355.

\bibitem{Hor11} Horvath, S. (2011). Weighted network analysis: applications in genomics and systems biology. Springer Science \& Business Media.

\bibitem{Ka53} Katz, L. (1953). A new status index derived from sociometric analysis. {\it Psychometrika} 18(1), 39-43.

\bibitem{LaFrHe10} Landherr, A., Friedl, B., \& Heidemann, J. (2010). A critical review of centrality measures in social networks. Wirtschaftsinformatik, 52, 367-382.



\bibitem{LiHaHu13}  Lindelauf, R. H., Hamers, H. J., \& Husslage, B. G. M. (2013). Cooperative game theoretic centrality analysis of terrorist networks: The cases of jemaah islamiyah and al qaeda. {\it European Journal of Operational Research}, 229(1), 230-238.

\bibitem{MaOrPo20}
 Manuel, C., Ortega, E., del Pozo, M. (2020). Marginality and Myerson values. {\it European Journal of Operational Research}, 284(1), 301-312.

 \bibitem{MaOrPo22} Manuel, C., Ortega, E., del Pozo, M. (2022). Marginality and the position value. {\it TOP}, 1-16.
 
\bibitem{Me88} Meessen, R. {\it Communication Games} (in Dutch). Master's Thesis, Department of Mathematics, University of Nijmegen, The Netherlands, 1988.

\bibitem{MuDiBo16} Musegaas, M., Dietzenbacher, B. J., \& Borm, P. E. (2016). On Shapley ratings in brain networks. {\it Frontiers in neuroinformatics}, 10, 51.


\bibitem{My77} Myerson, R. (1977). Graphs and cooperation in games. {\it Mathematics of Operations Research} 2, 225-229.

\bibitem{NaNa10} Narayanam, R. and Narahari, Y. (2010). A Shapley value-based approach to discover influential nodes in social networks. {\it IEEE Transactions on Automation Science and Engineering}, 8(1), 130-147.

\bibitem{Nw05} Newman, M. E. (2005). A measure of betweenness centrality based on random walks. {\it  Social Networks} 27(1), 39-54.

\bibitem{Ow86} Owen, G. (1986). Values of graph-restricted games. {\it SIAM Journal on Algebraic Discrete Methods}, 7(2), 210-220.

\bibitem{PoMaGoOw11} del Pozo, M., Manuel, C., Gonz\'{a}lez-Arang\"{u}ena, E., \& Owen, G. (2011). Centrality in directed social networks. A game theoretic approach. {\it Social Networks}, 33(3), 191-200.

\bibitem{Sa66} Sabidussi, G. (1966). The centrality index of a graph. {\it Psychometrika} 31(4), 581-603.

\bibitem{Sh53} Shapley, L. S. (1953). A Value for n-Person Games. In: {\it  Contributions to the Theory of Games II (Annals of Mathematics Studies 28)}; H. W. Kuhn and A. W. Tucker (eds.). Princeton University Press, 307-317. 

\bibitem{SkMiRa18} Skibski, O., Michalak, T. P., \& Rahwan, T. (2018). Axiomatic characterization of game-theoretic centrality. {\it Journal of Artificial Intelligence Research}, 62, 33-68.

\bibitem{SkRaMiYo19} Skibski, O., Rahwan, T., Michalak, T.P., Yokoo, M. (2019). Attachment centrality: Measure for connectivity in networks. {\it Artificial Intelligence} 274, 151-179.

\bibitem{Sl05}  Slikker, M. (2005). A characterization of the position value. {\it International Journal of Game Theory} 33 (4), 505-514.

\bibitem{TaMiRaWo17} Tarkowski, M. K., Michalak, T. P., Rahwan, T. and Wooldridge, M. (2017). Game-theoretic network centrality: A review. arXiv preprint arXiv:1801.00218.


\bibitem{We88} Weber, R.J. (1988). Probabilistic values for games. In: {\it The Shapley Value: Essays in Honor of Lloyd S. Shapley}; Roth, A (Ed.). Cambridge University Press, pp. 101-119.

\bibitem{ZuAkBoLiPr18} Zubiaga, A., Aker, A., Bontcheva, K., Liakata, M., \& Procter, R. (2018). Detection and resolution of rumours in social media: A survey. {\it ACM Computing Surveys (CSUR)}, 51(2), 1-36.

\end{thebibliography}
\end{document}